\documentclass[12pt]{iopart}

 \usepackage{epsfig}

\usepackage{amssymb}

\usepackage{multirow}

\begin{document}

\title{On the rise of proton-proton cross-sections at high energies}

\author{D A Fagundes, M J Menon and P V R G Silva}

\address{Universidade Estadual de Campinas - UNICAMP, Instituto de F\'{\i}sica Gleb Wataghin, 13083-859 Campinas, SP, Brazil}

\eads{\mailto{fagundes@ifi.unicamp.br}, \mailto{menon@ifi.unicamp.br}, \mailto{precchia@ifi.unicamp.br}}

\begin{abstract}
The rise of the total, elastic and inelastic hadronic cross sections at high energies is investigated
by means of an analytical parametrization, with the exponent of the leading logarithm 
contribution as a free fit parameter. Using derivative dispersion relations
with one subtraction,
two different fits to proton-proton and antiproton-proton total cross section
and rho parameter data are developed, reproducing well the experimental information
in the energy region 5 GeV - 7 TeV.
The parametrization for the total cross sections is then extended to fit the elastic (integrated)
cross section data in the same energy region, with satisfactory results.
From these empirical results we extract the energy dependence of several physical
quantities: inelastic cross section, ratios elastic/total,
inelastic/total cross sections, ratio total-cross-section/elastic-slope, elastic slope 
and optical point. All data, fitted and predicted, are quite well described.
We find a statistically consistent solution indicating:
(1) an increase of the hadronic cross sections with the energy faster than the 
log-squared bound by Froissart and Martin; (2) asymptotic limits 1/3 and 2/3 for the ratios elastic/total
and inelastic/total cross sections, respectively, a result in agreement with unitarity.
These indications corroborate recent theoretical arguments by Ya. I. Azimov on the rise of the total cross section.
\end{abstract}


\vspace{2.0cm}

\centerline{\textit{J. Phys. G: Nucl. Part. Phys.} (2013)}

\maketitle


\section{Introduction}\label{sec:1}

The rise of the hadron-hadron total, elastic and inelastic cross sections 
at high energies is an experimental fact. However, the exact energy dependence
involved and a widely accepted theoretical 
explanation for this increase have been open problems for a long time.
At present, these aspects constitute one
of the greatest challenges for QCD, the gauge field theory of the strong interactions.
The total cross section is connected with the forward \textit{elastic amplitude}
(optical theorem)
and therefore, perturbative techniques cannot access the large distance phenomena that permeate
the elastic channel. On the other hand, non-perturbative approaches are not yet able to evaluate
\textit{soft scattering states} from first principles. As a consequence, the 
dependence of the total hadronic
cross section with the energy cannot yet be directly predicted by QCD.

In the phenomenological context, a wide variety of models
present good descriptions of the available data. However, since they
are characterized by distinct physical pictures  \cite{pred,land,dremin,kaspar,fiore,matthiae},
not exclusively related to QCD, a global, unified and well accepted
approach is still missing.

In the theoretical context, formal results from axiomatic local quantum field theory,
general principles and further arguments, 
state that the total cross section 
cannot grow with the energy faster than log-squared
\cite{froissart,martin,lukmar}, a result 
also extended to the inelastic cross section \cite{martinin1,martinin2,wmrs}.
However, as recently discussed by Azimov \cite{azimov1,azimov2,azimov3}, it is not obvious if
these derivations can be directly applied to hadronic processes (QCD) and formal arguments suggest
that the total cross section could grow faster than log-squared. Moreover, recent result 
from lattice QCD indicates an universal asymptotic logarithm-squared dependence for the 
hadronic total cross section,
once some specific assumptions are made \cite{meg}. That is, this lattice QCD result
does not represent a unique or exclusive solution.

In the experimental context, new results from the LHC
and new estimates from the Pierre Auger Observatory for the
proton-proton ($pp$) total cross sections are expected to shed light on the subject.
In particular, the first result for the $pp$ total cross section at 7 TeV (LHC) has been obtained 
by the TOTEM Collaboration, through a \textit{luminosity-dependent measurement} \cite{totemsigma}.
Although, in principle, this is an experimental result considered rather
conservative due to agreement with some model predictions, further investigations
brought out  some unexpected aspects. In fact, once included in amplitude analysis
this point can not be adequately described by standard parametrization with leading
log-squared dependence on the energy, as first discussed
by us in \cite{fms11} (hereafter referred to as FMS) and recently (2012) also indicated 
by the Particle Data Group \cite{pdg12}. In this respect, 
see Figure 46.10 in [19] (reviewed version on line): within the 
corresponding uncertainties, the $\ln^2(s)$ fit result lies 
below the high-precision TOTEM datum; see also "Note Added 
During Revision" in \cite{fms11}.

The main ingredient in the FMS analysis has been the use of 
an analytical parametrization introduced by
Amaldi \textit{et al} in the seventies \cite{amaldi} and also used by the UA4/2 Collaboration,
in the nineties \cite{ua42}. This parametrization is characterized by power contributions at low
energies (reggeons) and a leading contribution at high energies (pomeron), parametrized by 
a power law in the logarithm of the energy with the exponent as a free real parameter.
Detailed analysis, with different methods and fit variants to total cross section data only,
from  $pp$ and antiproton-proton ($\bar{p}p$) scattering above 5 GeV,
led to the conclusion that the real exponent
in the logarithmic term is statistically consistent with 2 if the TOTEM datum is
not included in the analysis, but above 2 (around 2.2 - 2.3) if this point takes part
in the fitted data-set. 

In this work we extend the FMS analysis on $pp$ and $\bar{p}p$ scattering above 5 GeV in several aspects, 
all of them including the result of the TOTEM Collaboration at 7 TeV.

\begin{enumerate}

\item[1.] 
As discussed in \cite{fms11}, since the analytical parametrization is nonlinear in several parameters,
the data reductions demand start (feedback) values 
for the fit parameters \cite{bev} and depending on the choices the fit result may be different. 
Here, two different choices have been considered, 
one from the FMS analysis \cite{fms11} and another one based on the recent results from the PDG \cite{pdg12}.

\item[2.] 
Using derivative dispersion relations (DDR),
we connect the total cross section, $\sigma_{tot}$, with the $\rho$ parameter (ratio between the real 
and imaginary parts of the forward amplitude). With this analytical formalism, we first present
the predictions for $\rho(s)$ from individual fits to $\sigma_{tot}$ data and after that
we develop novel simultaneous fits to $\sigma_{tot}$ and $\rho$ data. In both cases
we discuss in some detail the important role of the subtraction constant,
embodied in the dispersion relations.

\item[3.] 
As commented on above, the log-squared bound,
originally stated for the total cross section by Froissart-Martin-Lukaszuk
\cite{froissart,martin,lukmar},  
\begin{equation*}
\sigma_{tot}(s) < \frac{\pi}{m_{\pi}^2}\, \ln^2 \left(\frac{s}{s_r}\right),
\end{equation*}
where $s$ is the  center-of-mass energy squared, $s_r$ a constant, $m_{\pi}$
the pion mass,
has been recently
extended to the inelastic cross section by Martin-Wu-Roy-Singh \cite{martinin1,martinin2,wmrs},
\begin{equation*}
\sigma_{inel}(s) <  \frac{1}{4} \frac{\pi}{m_{\pi}^2}\, \ln^2 \left(\frac{s}{s_r}\right).
\end{equation*}
Therefore, from s-channel unitarity, a similar bound is expected for
the elastic (integrated) cross section, $\sigma_{el}(s)$.
Based on the connection between the total cross section and the forward
\textit{elastic scattering amplitude} (optical theorem) and since the FMS analysis of the total cross section has 
indicated a power larger than 2, we address here the extension of the same parametrization to the
\textit{elastic cross section}. Namely, with start (feedback) values of the simultaneous fits
to $\sigma_{tot}$ and $\rho$ data we develop new empirical fits  to $\sigma_{el}$ data,
with the same analytical parametrization.

\item[4.] 
With the results from the empirical fits for $\sigma_{tot}(s)$, $\rho(s)$ and  $\sigma_{el}(s)$, we  present
\textit{predictions} for the energy dependence of 
several physical quantities:
$\sigma_{inel}$, the ratios 
$\sigma_{el}/\sigma_{tot}$,
$\sigma_{inel}/\sigma_{tot}$,
the optical point, as well as a result for the
ratio between $\sigma_{tot}$
and the elastic slope $B$ and from that for the proper slope $B$.

\item[5.] 
Here, all physical quantities that are compared with experimental data (fitted and predicted
quantities), are displayed with uncertainty regions and are
evaluated by means
of standard (analytical) error propagation from the free fit parameters \cite{bev}.

\end{enumerate}

All the experimental data considered (fitted and predicted) are quite
well described.
Our analysis leads to solutions that favor an  increase of the total 
and elastic cross sections faster than the log-squared bound. 
In that case, the violation of the bound does not imply in violation of unitarity,
as recently discussed by Azimov.
Indeed, we obtain that, asymptotically, $\sigma_{el}/\sigma_{tot} \rightarrow$ 1/3
and $\sigma_{inel}/\sigma_{tot} \rightarrow$ 2/3.
However, although statistically consistent and corresponding to quite
good descriptions of the existing data,
our results do not constitute unique solutions, but possible solutions
for the rise of the hadronic cross sections.

The manuscript is organized as follows. In \sref{sec:2} we treat the
fits to $\sigma_{tot}$ and $\rho$ data: the analytical
parametrization for $\sigma_{tot}$ is introduced and the analytical connection with 
$\rho$ by means of dispersion relations is presented and discussed, with emphasis on
the role of the subtraction constant.
In \sref{sec:3} the extension of the parametrization to  elastic cross sections data is presented. 
In \sref{sec:4} we display the predictions for the inelastic cross section, the ratios involving cross sections,
the optical point, the elastic slope and compare with the corresponding experimental data,
followed by discussions on each obtained result. 
Our conclusions and some critical remarks 
are the contents of \sref{sec:5}.

\section{Total Cross Section and the Rho Parameter}\label{sec:2}

At high energies, in terms of the \textit{elastic} scattering amplitude $F$, the total cross section and the 
$\rho$ parameter in the forward direction can be expressed by
\begin{eqnarray}
\sigma_{tot}(s) = \frac{\mathrm{Im}\, F(s, t=0)}{s}
\qquad \mathrm{(Optical\ theorem)}
\end{eqnarray}

\begin{eqnarray}
\rho(s) = \frac{\mathrm{Re}\, F(s, t=0)}{\mathrm{Im}\, F(s, t=0)},
\end{eqnarray}
where $t$ is the four momentum transfer squared.

\subsection{Experimental Data}\label{subsec:2.1}

Our interest here is the investigation of the high and asymptotic
energy region, associated with particle-particle and antiparticle-particle
scattering.
For that reason we shall consider
only elastic collisions with the \textit{highest energy interval in terms of available data}, namely
$pp$ and $\bar{p}p$ scattering.
With this restrictive choice we do not take into account 
data from other reactions, available only in the regions
of intermediate and low energies or any constraint dictated by a
supposed universal behavior.

The input data set for fits concerns \textit{only accelerator data}, covering
the region from 5 GeV up to 7 TeV. However, estimations
of the $pp$ total cross section from cosmic-ray experiments will be
displayed as illustrative results. The reason why we do not include this information
in the fits is the model-dependence
involved in the extraction of the proton-proton total cross section from proton-air production cross sections
\cite{engel1,engel2,engel3,alm03}.
The accelerator data on $\sigma_{tot}$ and $\rho$ from $pp$ and $\bar{p}p$ scattering have been extracted from
\cite{totemsigma} and the PDG database \cite{pdgsite}, without any kind of data selection or sieve procedure.
The  estimations of the $pp$ total cross sections from cosmic-ray experiments
are from \cite{pdgsite,auger}. All these data and estimations are exactly the same 
as that used and displayed in \cite{fms11}.

\subsection{Analytical Parametrization}\label{subsec:2.2}

The analytical parametrization to be used here in the fits to $pp$ and $\bar{p}p$
total cross sections data (and in \sref{sec:3} to elastic cross sections data), 
is given by \cite{fms11,amaldi,ua42}

\begin{eqnarray}
\sigma^{pp}(s) =  
a_1\, \left[\frac{s}{s_l}\right]^{-b_1} - 
\, a_2\, \left[\frac{s}{s_l}\right]^{-b_2} +
\alpha\, + \beta\, \ln^\gamma \left(\frac{s}{s_h}\right),
\end{eqnarray}

\begin{eqnarray}
\sigma^{\bar{p}p}(s) =  
a_1\, \left[\frac{s}{s_l}\right]^{-b_1} +
\, a_2\, \left[\frac{s}{s_l}\right]^{-b_2} +
\alpha\, + \beta\, \ln^\gamma \left(\frac{s}{s_h}\right),
\end{eqnarray}
where $s_l$ = 1 GeV$^2$ (fixed), $a_1$, $b_1$, $a_2$, $b_2$ (low energies),
$\alpha$, $\beta$, $\gamma$ and $s_h$ (high energies) are, 
in general, free real fit parameters.

For the case of total cross section all the
parameters above have specific physical interpretations in the context
of the Regge-Gribov theory. They are associated with reggeon exchanges (first two terms)
and pomeron exchanges (last two terms) at
low and high energy regions, respectively \cite{fms11,compete1}.
In the case of the elastic cross section we consider this parametrization
as an empirical ansatz or as a representation. 

\subsection{Dispersion Relations and the Subtraction Constant}\label{subsec:2.3}

Amplitude analyses of the total cross section rising usually include
the information on the $\rho$ parameter through fits to both
quantities \cite{compete1,compete2,ckk,lm,igi1,igi2,igi3,bh1,bh2,bh3}.
The connection between $\sigma_{tot}$ and $\rho$ is obtained by means of 
dispersion relations (integral and/or
derivatives forms), or the asymptotic prescriptions for crossing even and odd
amplitudes (Phragm\'en-Lindel\"off theorem) \cite{eden,bc}. 
In the high-energy region, of interest in this work, the dispersion relations
demand one subtraction \cite{disprel1,disprel2}, which means the addition of one more free parameter
in the simultaneous investigation of both quantities.

In the FMS analysis \cite{fms11} the fits have been developed only with
the $\sigma_{tot}$ data, without the inclusion of the $\rho$ information.
Here, using DDR with one subtraction,
we also treat simultaneous fits to both quantities.
To this end,
in this subsection, we shortly review some aspects related to  integral and DDR, 
with emphasis on the origin and role of the subtraction
constant. In the next subsections, we present our predictions for $\rho(s)$ from fits
to $\sigma_{tot}$ data and after that, we develop and discuss new simultaneous fits to $\rho$ and $\sigma_{tot}$
data.

\subsubsection{Integral and Derivative Dispersion Relations}\label{subsubsec:2.3.1}

Following the notation in \cite{alm03}, for $pp$ and $\overline{p}p$ scattering, analyticity and crossing symmetry
allow us to connect $\sigma_{tot}(s)$ and $\rho(s)$ through two compact and
symmetric formulas:  
\begin{eqnarray}
\rho ^{pp}(s) \sigma_{tot}^{pp} (s) =
E( \sigma_{+} ) + O( \sigma_{-} ),
\end{eqnarray}

\begin{eqnarray}
\rho ^{\bar{p}p}(s) \sigma_{tot}^{\bar{p}p} (s) =
E( \sigma_{+} ) - O( \sigma_{-} ),
\end{eqnarray}
where the even ($+$) and odd ($-$) cross section are related to the physical cross sections by

\begin{eqnarray}  
\sigma_{\pm}(s) = \frac{\sigma_{tot}^{pp} \pm 
\sigma_{tot}^{\bar{p}p}}{2},
\end{eqnarray} 
and $E( \sigma_{+} )$, $O( \sigma_{-} )$ are analytic transforms 
connecting the
real and imaginary parts of crossing even and odd scattering amplitudes, 
respectively (associated with $\sigma_{+}$ and $\sigma_{-})$.
These analyticity relations are usually expressed in an \textit{integral} 
form (Hilbert transform \cite{byron}) and
in the case of the forward direction the standard once-subtracted
\textit{integral dispersion
relations} (IDR), with poles removed, may be expressed by \cite{bc,disprel1,disprel2,acm}
 
\begin{eqnarray} 
E_{int}(\sigma_{+}) \equiv
\frac{K}{s} + \frac{2s}{\pi}\,P\int_{s_o}^{\infty} 
ds' \left[\frac{1}{s'^2-s^2}\right]  
\sigma_{+}(s'), 
\end{eqnarray}

\begin{eqnarray} 
O_{int}(\sigma_{-}) \equiv
  \frac{2}{\pi}\,P\int_{s_o}^{\infty} 
ds' \left[\frac{s'}{s'^2-s^2}\right]  
\sigma_{-}(s'),
\end{eqnarray}
where $K$ is the \textit{subtraction constant} and $P$ denotes principal value.
We recall that in the even case, IDR with one and two
subtractions are equal and in the odd case, those without subtraction and
with one subtraction are also equal \cite{bc}. Therefore, with
our parametrization of interest (3-4), the subtraction in the even part 
is adequate since, \textit{asymptotically}, from  equation (7),
we have:

\begin{eqnarray} 
\sigma_{+} \rightarrow \ln^{\gamma}(s)\ \ \mathrm{and} \ \ \sigma_{-} \rightarrow 0
\ \ \mathrm{as} \ \ \sqrt{s}  \rightarrow \infty.
\nonumber
\end{eqnarray}
We shall return to the subject of the subtraction constant in the next subsection.

On the other hand, for classes of functions of interest, IDR
can be replaced by derivative forms \cite{ddr1,ddr2,ddr3}, known as \textit{derivative dispersion relations}
(DDR),
which may be more
useful for some practical calculations.
In these formulas, differentiation with
respect to the logarithm of the energy occurs in the argument of a
trigonometric operator, as in the
standard form deduced by Bronzan, Kane, and Sukhatme 
\textit{in the high-energy approximation} ($s_0 \rightarrow 0$ in eqs. (8) and (9)) \cite{bks,am04}:

\begin{eqnarray} 
E_{der}(\sigma_{+}) \equiv
\frac{K}{s} + 
\tan\left[\frac{\pi}{2}\frac{d}{d\ln s} \right] 
\sigma_{+}(s), 
\end{eqnarray}

\begin{eqnarray} 
O_{der}(\sigma_{-}) \equiv
\tan\left[\frac{\pi}{2}\left( 1 + \frac{d}{d\ln s}\right) \right]
\sigma_{-}(s),
\end{eqnarray} 
corresponding, as before, to singly-subtracted DDR.

This completes the analytical approach: with an input parametrization for 
$\sigma^{pp}_{tot}(s)$, $\sigma^{\bar{p}p}_{tot}(s)$, eqs. (5), (6) and (7) allow, 
in principle, the determination of $\rho^{pp}(s)$ and $\rho^{\bar{p}p}(s)$ by
means of either the IDR, (8) and (9), or the DDR, 
(10) and (11) in the high-energy approximation.

In our case, with the $\ln^{\gamma}(s/s_h)$ term and
$\gamma$  a free real parameter, IDR 
demand numerical methods 
and therefore a nonanalytic approach for error propagation from the fit parameters.
The use of prescriptions seems to us unjustified in the region of
intermediate and low energy data (5 - 20 GeV), since they are asymptotic results \cite{eden}.
On the other hand, DDR allow an analytical approach and 
even the high-energy results, (10) and (11), can be applied in the low energy region,
an effect related to the presence of the subtraction constant $K$. Since this
effect plays a fundamental role in our analysis, let us
first recall the origin, meaning and role of this constant, with focus on its disadvantages and
\textit{advantages in the present case}.

\subsubsection{The Role of the Subtraction Constant}\label{subsubsec:2.3.2}

Subtractions are useful tools to deal with the convergence of IDR and they are essentially
based on a change of variable. In general grounds, suppose a given dispersion relation for a
complex function $f(z)$ and let $x_0$ be some point on the real axis, at which $f(z)$ is
analytic. The corresponding relation for $[f(z) - f(x_0)] / [z - x_0]$ leads to the
subtracted dispersion relation \cite{bc,byron,mmp}. Separating the real an imaginary parts of 
$f(z)$, the IDR for Re$ f(z)$ includes now the constant Re$ f(x_0)$, the subtraction constant.
Therefore, in a particular physical problem it is necessary to determine Re$ f(x_0)$ for
some $x_0$.
For scattering amplitudes $x_0 \rightarrow s = 0$ is assumed, so that for an
even amplitude ($F_{+}$), or cross section as in eq. (8), the subtraction constant reads
\begin{eqnarray} 
K = \mathrm{Re}\, F_{+}(s=0),
\nonumber
\end{eqnarray} 
meaning the analytic continuation of the even amplitude at $s=0$ \cite{bc}.

Therefore, the subtraction constant has a well justified mathematical origin,
related to the convergence of the integral. On the other hand, we cannot attribute to it
a (real) physical meaning or value, since it is associated with the nonphysical region.
However, in applying dispersion relations to connect $\sigma_{tot}$ and $\rho$
this constant appears explicitly and therefore must be taken into account. Note that
to assume $K=0$ is only a particular (and unjustified) choice, as any other.
In this respect two disadvantages are in order to be recalled.

1. As an unknown constant, it plays the role of one more free parameter. However,
the lack of a physical meaning contrasts with all the other
free parameters present in the analytical parametrization of the total cross section,
namely reggeons intercepts, reggeons strengths and leading pomeron contribution.

2. In data reductions, as a free parameter, its value affects the final values
of the others physical free parameters.
Moreover, it may seem that, once appearing in the form $K/s$ in equations (8) and (10),
its influence is limited to the low energy region. However, that is not true: typical (global)
fits are characterized by \textit{strong correlation among all free parameters}. That means 
$K$ may affect the fit results even at high and asymptotic energies, as already
discussed and demonstrated in \cite{alm03,am04,alm01}.

Summarizing, once introduced in the fit procedure, one may, in principle,  select and even 
anchor the outputs. That was one of the
reasons why simultaneous fits, including the $\rho$ information, were not considered in the FMS 
analysis (see however
Appendix A in \cite{fms11}).

On the other hand, \textit{in applying DDR deduced in the high-energy approximation}
(previous subsection), the subtraction constant develops a useful property, related
to the equivalence between DDR and IDR. These aspects have been already demonstrated and discussed
in \cite{alm03,am04,alm01} and in what follows we recall the main points.

As commented before,
the DDR deduced by Bronzan-Kane-Sukhatme (10-11) have been obtained
in the high-energy approximation, which means considering $s_0 \rightarrow 0$ in the
lower limit of the IDR (8) and (9). Since for $pp$ and $\bar{p}p$ scattering, 
$s_0 = 4m_p^2 \sim $ 3.5 GeV$^2$, it could be expected that, for a given parametrization, the results 
obtained by means of IDR with $s_0 = 4m_p^2$ fixed and through DDR are not the same.
That, however, is not always the case.
In \cite{am04} and references therein, the practical equivalence of IDR and DDR has
been investigated in the case of the Donnachie-Landshoff parametrization, with
both degenerate \cite{dl} and non-degenerate \cite{ckk,cmg} meson trajectories.
Several variants have been considered in simultaneous fits to $\sigma_{tot}$ and $\rho$ data
from $pp$ and $\bar{p}p$ scattering: 
(i) energy cutoffs at 5 and 10 GeV;
(ii) subtraction constant $K = 0$ and $K$ as a free fit parameter;
(iii) DDR and IDR with both $s_0 = 4m_p^2$ and $s_0 = 0$.
In all cases, once the subtraction constant is used as a free
parameter, the numerical results obtained through DDR (high-energy approximation) and IDR with fixed $s_0 = 4m_p^2$
(without the high-energy approximation)
are the same, up to three significant figures in the fit parameters and $\chi^2$/DOF
(degrees of freedom).
This effect is a consequence of the absorption of the high-energy approximation
in an \textit{effective} subtraction constant. Specifically, as analytically
demonstrated in Sect. 4.4 of \cite{am04}, the coefficient $\Delta$ of the series expansion associated
with the integration from $s'= 0$ to $s'= 4m_p^2$ is absorbed by the subtraction constant,
since the series can be put in the form

\begin{eqnarray} 
[ K + \Delta]\, \frac{1}{s} + \mathcal{O}\left(\frac{1}{s^2}\right).
\nonumber
\end{eqnarray} 
This effect of absorption of the high-energy approximation by the subtraction constant
has also been confirmed by other authors \cite{cudell1,cmsl}.
Therefore, despite the disadvantages of the subtraction constant, it can have here a 
useful practical or \textit{pragmatic} role as a free fit parameter in the DDR: \textit{it leads to the same
result that could be obtained through the IDR without the high-energy approximation}.

We recall that, for classes of functions of interest in high-energy elastic scattering,
DDR can be analytically extended down to 4 - 5 GeV \cite{cudell1,cudell2} or even below 
and for the whole energy interval (above the
physical threshold), either 
in the form of a double infinity series, as first deduced by \'Avila and Menon \cite{am1,am2}, or 
in the form of a single series, as demonstrated by Ferreira and Sesma \cite{fs}. 
However it seems to us simpler here to consider the
pragmatic role of the subtraction constant, since as explained above,
with this  constant as a free parameter the results obtained with the DDR (10) and (11)
(high-energy approximation) are the same as those obtained through the IDR (8) and (9) (without the high-energy
approximation). Therefore, let us focus on the DDR (10) and (11) with $K$ as a free fit parameter.

\subsubsection{Analytical results with Derivative Dispersion Relations}\label{subsubsec:2.3.3}

The applicability of DDR in amplitude analysis
has been critically reviewed by \'Avila and Menon \cite{am04}, in particular the replacement
of the trigonometric operators in (10) and (11) by the corresponding series
(see also \cite{alm03}).
In practice, with the presence of the $\ln^{\gamma}(s/s_h)$ term,
the derivative transforms (10) and (11) can be evaluated through the operational expansion,
introduced by Kang and Nicolescu \cite{kn} (see also section 3 in \cite{am04}):

\begin{eqnarray}\fl 
E_{der}(\sigma_{+}) =\frac{K}{s} +
\left[ \frac{\pi}{2} \frac{d}{d\ln s} + 
\frac{1}{3} \left(\frac{\pi}{2}\frac{d}{d \ln s}\right)^3 + \frac{2}{15} \left(\frac{\pi}{2}\frac{d}{d \ln s}\right)^5 + \dots \right] \sigma_{+}(s),
\end{eqnarray}

\begin{eqnarray} \fl
O_{der}(\sigma_{-}) =
- \int \left\{ \frac{d}{d\ln s} \left[\cot \left( \frac{\pi}{2} 
\frac{d}{d\ln s} \right)\right] \sigma_{-}(s) \right\} d\ln s \nonumber \\
\!\!\!\!\!\!\!\!\!\!\!\! = -\frac{2}{\pi}\int \left\{ \left[ 1 - \frac{1}{3} \left(\frac{\pi}{2}\frac{d}{d \ln
s}\right)^2 -  \frac{1}{45} \left(\frac{\pi}{2}\frac{d}{d \ln s}\right)^4
 -\dots \right] \sigma_{-}(s) \right\} d \ln s.
\end{eqnarray}

With parametrization (3-4) as input and from (7), the evaluation of the power terms (low energy contribution)
results in closed forms (sum of the series). For the logarithm term (leading contribution)
a fast convergence is already obtained up to third order, since as we shall show $\gamma \sim 2.2 - 2.4$.
In this case we obtain \cite{fms11}

\begin{eqnarray}
E_{der}(\sigma_{+}) = & \frac{K}{s} 
- a_1 \tan \left( \frac{\pi b_1}{2}\right) \left[\frac{s}{s_l}\right]^{-b_1}\!\!\!\! +
\mathcal{A} \ln^{\gamma - 1} \left(\frac{s}{s_h}\right) \nonumber \\
& + \mathcal{B} \ln^{\gamma - 3} \left(\frac{s}{s_h}\right) +
\mathcal{C} \ln^{\gamma - 5} \left(\frac{s}{s_h}\right),
\end{eqnarray} 
where
\begin{eqnarray} 
\mathcal{A} &=& \frac{\pi}{2} \, \beta\, \gamma,  \qquad \qquad
\mathcal{B} = \frac{1}{3} \left[\frac{\pi}{2}\right]^3 \, \beta\, \gamma\, [\gamma - 1][ \gamma - 2], \nonumber \\ 
                                                                 \nonumber \\
\mathcal{C} &=& \frac{2}{15} \left[\frac{\pi}{2}\right]^5 \, \beta\, \gamma\, [\gamma - 1][ \gamma - 2]
[\gamma - 3][ \gamma - 4]
\end{eqnarray} 
and for the odd contribution,

\begin{eqnarray} 
O_{der}(\sigma_{-}) =
-\, a_2\, \cot \left(\frac{\pi\, b_2}{2}\right) \left[\frac{s}{s_l}\right]^{-b_2}.
\end{eqnarray} 

With the above results ((14),(15),(16)) and parametrization (3)-(4) for $\sigma^{pp}_{tot}$ and $\sigma^{\bar{p}p}_{tot}$, 
equations (5) and (6) lead to the analytical expressions for $\rho ^{pp}(s)$ and $\rho ^{\bar{p}p}(s)$:

\begin{eqnarray}\fl
\rho ^{pp}(s) = \frac{1}{\sigma_{tot}^{pp}(s)}
\left\{ \frac{K}{s} 
- a_1\, \tan \left( \frac{\pi b_1}{2}\right) \left[\frac{s}{s_l}\right]^{-b_1} +
\mathcal{A} \ln^{\gamma - 1} \left(\frac{s}{s_h}\right) \nonumber  \right. \\ 
\qquad +  \left.\mathcal{B} \ln^{\gamma - 3} \left(\frac{s}{s_h}\right) 
+\mathcal{C} \ln^{\gamma - 5} \left(\frac{s}{s_h}\right) 
-  a_2\cot \left(\frac{\pi b_2}{2}\right) \left[\frac{s}{s_l}\right]^{-b_2} \right\}.
\end{eqnarray}
and
\begin{eqnarray}\fl
\rho ^{\bar{p}p}(s) = \frac{1}{\sigma_{tot}^{\bar{p}p}(s)} 
\left\{\frac{K}{s} 
- a_1\, \tan \left( \frac{\pi b_1}{2}\right) \left[\frac{s}{s_l}\right]^{-b_1} +
\mathcal{A} \ln^{\gamma - 1} \left(\frac{s}{s_h}\right)\nonumber \right. \\
\qquad +\left.
\mathcal{B} \ln^{\gamma - 3} \left(\frac{s}{s_h}\right)  
+\mathcal{C} \ln^{\gamma - 5} \left(\frac{s}{s_h}\right)
+ a_2 \cot \left(\frac{\pi b_2}{2}\right) \left[\frac{s}{s_l}\right]^{-b_2} \right\},
\end{eqnarray}
with $\mathcal{A}$, $\mathcal{B}$ and $\mathcal{C}$ given by eq. (15).
Note that the analytical results imply that as $s \rightarrow \infty$: 
\begin{eqnarray} 
\rho \ \propto \ \frac{1}{\ln s} \ \rightarrow \ 0,
\nonumber
\end{eqnarray}
which is in agreement with the rigorous asymptotic result by Khuri and Kinoshita \cite{kk}.

\subsection{Fits and Results}\label{subsec:2.4}

In this subsection, after discussing general aspects of the fit procedures and methodology,
we present the results, through eqs. (17) and (18), of individual fits to $\sigma_{tot}$ data with corresponding
predictions for the $\rho(s)$ and those from simultaneous fits to $\sigma_{tot}$ and
$\rho(s)$ data. A detailed discussion of all the results obtained is presented in the
next subsection.

\subsubsection{Fit Procedures and Methodology.}\label{subsubsec:2.4.1}

The nonlinearity of the fit (constants $b_1$, $b_2$, $\gamma$ and $s_h$ in equations (3)-(4)) demands
a choice of the initial (feedback) values of the free parameters \cite{bev}. Here we shall address 
two different choices, denoted and explained in what follows.

\begin{description}

\item[\textbf{Fit 1}.]
In the FMS analysis, based only on the total cross section data, two ensembles, two methods and 
six variants have been tested for different feedback values and different fixed and free parameters.
Here we shall select as a representative
result with $\gamma > $ 2 the one denoted in \cite{fms11} as Ensemble $\sqrt{s}_{\mathrm{max}}$ = 7 TeV, method 2 
and variant 5
(table 4 and figure 6 in \cite{fms11}). The reason for this choice is five-fold: 

(i) $\chi^2/$DOF 
closest to 1; 

(ii) smaller
relative error in the exponent $\gamma$; 

(iii) result closest to the TOTEM point at 7 TeV;

(iv) the numerical value of the exponent $\gamma$ does not correspond to the largest or the smallest 
value obtained
in the analysis figure 8 in \cite{fms11}). 

(v) the reggeon intercepts, $b_1$ and $b_2$, are fixed to 0.5, so that we can investigate
the effects of this assumption (compared with fit 2 discussed below, in which the intercepts
are not fixed).

\item[\textbf{Fit 2}.]
The 2012 PDG analysis \cite{pdg12} is based on
the highest rank parametrization result by the COMPETE Collaboration,
introduced ten years ago \cite{compete1, compete2}. The functional form can be seen as 
a particular case of parametrization (3-4) in which
the exponent $\gamma$ is fixed to 2. The analysis by PDG includes different reactions and tests
on the universatility. Since this work corresponds to another fit result, 
with a kind of ``conservative character'' (namely $\gamma$ = 2), we have considered
the values of the parameters they have obtained as start values for fits with parametrization (3-4). 
We note that in doing so we are applying the parametrization to a subset of the data
analyzed by the PDG and by the COMPETE Collaboration.

\end{description}

The data reductions have been performed with the objects of the class TMinuit of ROOT Framework 
\cite{root} and all results correspond only to the cases of full convergence. As tests of goodness of fit
we consider the $\chi^2$ per DOF, with the $\chi^2$ directly obtained from
the MINUIT Code. 
In all fits we have adopted 
the Confidence Level (CL) of $\approx$ 68 $\%$ (one standard deviation),
which means that the projection of the $\chi^{2}$ distribution in $(N+1)$-dimensional space 
($N =$ number of free fit parameters) contains 68 $\%$ of probability \cite{bev}.

In the MINUIT Code, the correlation matrix gives a measure of the correlation between
each pair of free parameters, with numerical limits $\pm$ 1 (full correlation) and 0
(no correlation) \cite{bev} (we shall return to this point in what follows).
The error matrix provides the variances and covariances associated with each free parameter.
This information is used in the analytic evaluation of the uncertainty regions in the fitted
and extracted (predicted) physical quantities through standard error propagation procedures \cite{bev}.

In what follows we shall address some of the points raised on the role of the
subtraction constant in \sref{subsubsec:2.3.2}, now with the results related
to parametrization (3-4), fits 1 and 2 and the derivative dispersion results (17) and (18).
First we treat the predictions for $\rho(s)$ from fits to $\sigma_{tot}$ data and after that, simultaneous fits
to both $\sigma_{tot}$ and $\rho$ data. 
As explained, the consistent applicability of the DDR demands the subtraction constant $K$
as a free fit parameter. However, in order to illustrate some aspects of this constant
we shall consider two variants in all cases: $K$ = 0 and $K$ as a free fit parameter.

\subsubsection{Individual Fits to Total Cross Section Data and Predictions
for the Rho Parameter}\label{subsubsec:2.4.2}

$\bullet$ \textbf{Total Cross Section}

\vspace{0.3cm}

In the case of fit 1, the values of the parameters obtained in \cite{fms11},  
through parametrization (3-4) and fit to only the total cross section data,
are displayed in the second column of \tref{tab:1} ($\sigma_{tot}$ data column),
with the corresponding statistical information on the data reduction.
The dependences of the cross sections on energy are shown in \fref{fig:1}, including uncertainty regions from error
propagation, together with the experimental data analyzed (accelerator) and estimations of $\sigma_{tot}^{pp}$
from cosmic-ray experiments.

In the case of fit 2 to  total cross section data, the feedback values of the parameters and the fit results are displayed in the 
fifth and sixth columns of \tref{tab:1}, respectively ($\sigma_{tot}$ data columns).
In this case, the parameter $s_M$ that appears associated with the power contributions 
in the PDG analysis, has been included in the reggeon strengths $a_1$ and $a_2$. We distinguish however
the scale factor in the logarithm contribution ($s_h$). 
The results for the total cross section, including uncertainty regions, and experimental
information are shown in \fref{fig:2}.

\vspace{0.3cm}

$\bullet$ \textbf{Predictions for $\rho(s)$}

\vspace{0.3cm}

From eqs. (17) and (18), any  prediction for $\rho ^{pp}(s)$ and $\rho ^{\bar{p}p}(s)$ 
from the total cross section fit depends on the value of the subtraction constant.
As commented before, we shall consider the two variants $K$ = 0 and $K$ as a free fit
parameter to the $\rho$ data.
Specifically, we use as input the fixed values of the parameters for the total
cross section (\tref{tab:1}), with $s_l$ = 1 GeV$^2$. For $K=0$ (fixed) we obtain the direct 
prediction for $\rho(s)$. In the
case of $K$ as a free parameter we use $K=0$ as start value and fit
only the $\rho$ data with this parameter, by fixing all the other
parameters (from the total cross sections). In this case the first run with the MINUIT Code allows us to obtain
the corresponding $\chi^2$/DOF for $K=0$.
This statistical information and the values of $K$ as fit parameter are displayed in
the third and fourth columns of \tref{tab:1} with fit 1 and
seventh and eighth column of \tref{tab:1} with fit 2 ($\rho$ data columns).
The dependences of $\rho$ on energy,
including uncertainty regions from error
propagation and the experimental data,
 are shown in \fref{fig:1} with fit 1 and \fref{fig:2} with
fit 2 (both for $K$ = 0 and $K$ as a free fit parameter).

\subsubsection{Simultaneous Fits to Total Cross Section and Rho Data}\label{subsubsec:2.4.3}

For simultaneous fits to $\sigma_{tot}$ and $\rho$ data, we have used as feedback the values of the
parameters from fit 1 and fit 2 to $\sigma_{tot}$, displayed in \tref{tab:1} (second and sixth columns).
Here, we also consider the two
variants: $K = 0$ and  $K$ as a free fit parameter (in this case, with $K = 0$ as start value).
The fit results and statistical information are displayed in \tref{tab:2}. 
The results for $\sigma_{tot}(s)$ and $\rho(s)$ from fit 1 are shown in
figures \ref{fig:3} ($K$ = 0) and \ref{fig:4} ($K$ as free parameter) and from fit 2 in 
figures \ref{fig:5} ($K$ = 0) and \ref{fig:6} ($K$ as free parameter).

\fulltable{\label{tab:1}Results of the individual fits to $\sigma_{tot}$ data and the predictions
for the $\rho$ parameter, with the subtraction constant $K=$ 0  and $K$ as a free
parameters and fits 1 and 2 (see text).
The results denoted fit 1 (second, third and fourth
columns) are those obtained in \cite{fms11}. In fit 2 we use 
as initial values of the free parameters the central 
values from the PDG 2012 result \cite{pdg12} (fifth column)
and our fit results are displayed in the sixth, seventh
and eighth columns.
Also included are the statistical
information on the data reductions: degrees of freedom (DOF) and reduced $\chi^2$. 
The parameters $a_1$, $a_2$, $\alpha$ and $\beta$
are in mb, $s_h$ in GeV$^{2}$ and $b_1$, $b_2$, $\gamma$ and $K$ are dimensionless.
In all cases, $s_l$ = 1.0 GeV$^2$ (fixed) in parametrization (3-4).}
\br
            & \centre{3}{Fit 1 \cite{fms11}:}         & \centre{4}{Fit 2:} \\
Ensemble:   &\;\;$\sigma_{tot}$ data  & \centre{2}{$\rho$ data} & \centre{2}{$\sigma_{tot}$ data}  &\centre{2}{$\rho$ data} \\
            &                   &$K=$0  & \;\; $K$ free & \;\;Initial values & \;\;Fit results \cite{pdg12}     &\;\;$K=$0 & \;\;$K$ free   \\\mr
$a_1$       &\;\;56.5$\pm$1.1                     & $\:-$  & $\:-$                 &\;\;46.06         &\;\;60.2$\pm$1.5                 & $\:-$   &  $\:-$\\
$b_1$       &\;\;0.5 (fixed)                      & $\:-$  & $\:-$                 &\;\;0.462         &\;\;0.4850$\pm$0.0079            & $\:-$   & $\:-$\\
$a_2$       &\;\;27.70$\pm$0.28                   & $\:-$  & $\:-$                 &\;\;34.0          &\;\;33.1 $\pm$ 1.7               & $\:-$   &  $\:-$\\
$b_2$       &\;\;0.5 (fixed)                      & $\:-$  & $\:-$                 &\;\;0.550         &\;\;0.540 $\pm$ 0.015            & $\:-$   &  $\:-$\\
$\alpha$    &\;\;33.65$\pm$0.22                   & $\:-$  & $\:-$                 &\;\;34.71         &\;\;31.52 $\pm$ 0.22             & $\:-$   &  $\:-$\\
$\beta$     &\;\;0.1301$\pm$0.0086                & $\:-$  & $\:-$                 &\;\;0.265         &\;\;0.0575$\pm$ 0.0025           & $\:-$   &  $\:-$\\
\boldmath $\gamma$ &\;\;\textbf{2.213$\pm$0.024}  & $\:-$  & $\:-$                 &\;\;2.0           &\;\;\textbf{2.422 $\pm$ 0.016}   & $\:-$   &  $\:-$\\
$s_h$       &\;\;3.90$\pm$0.52                    & $\:-$  &  $\:-$                &\;\; 16.2         &\;\;0.566 $\pm$ 0.065            & $\:-$   & $\:-$\\
$K$         &\; $\:-$                             & \;0    & 44.8$\pm$4.9          &\;$\:-$           &\; $\:-$                       & 0       &38.5$\pm$4.9\\ \mr
DOF         &\;\; 158                             & 76     & 75                    &\;$\:-$         &\;\;156                          & 76      & 75\\
$\chi^2/$DOF&\;\; 0.967                           &2.67    & 1.58                  &\;$\:-$         &\;\;0.934                       &2.28     & 1.48\\
\br
\endfulltable

\fulltable{\label{tab:2}Results of the simultaneous fits to $\sigma_{tot}$ and $\rho$ data,
with feedback values from fit 1 and fit 2 (\tref{tab:1}), for 
$K = 0$ and  $K$ as a free fit parameter. Same legend as \tref{tab:1}.}
\br
             & \centre{2}{From Fit 1:}          & \centre{2}{From Fit 2:}            \\
             & $K = 0$          & $K$ free  & $K = 0$           & $K$ free   \\\mr
$a_1$       &48.58 $\pm$ 0.72 &51.5 $\pm$ 4.3   &65.99 $\pm$ 0.55     &59.8 $\pm$ 1.3 \\
$b_1$       &0.5 (fixed)    &0.5 (fixed)    &0.2758 $\pm$ 0.0026  &0.4541 $\pm$ 0.0069 \\
$a_2$       &26.97 $\pm$ 0.27 &27.63 $\pm$ 0.28 &34.5 $\pm$ 1.5       &34.1 $\pm$ 1.6 \\
$b_2$       &0.5 (fixed)    &0.5 (fixed)    &0.5511 $\pm$ 0.0094  &0.547 $\pm$ 0.010 \\
$\alpha$    &35.41 $\pm$ 0.15 &35.16 $\pm$ 0.89 &5.59 $\pm$ 0.24      &29.78 $\pm$ 0.22 \\
$\beta$     &0.264 $\pm$ 0.041&0.32 $\pm$ 0.11  &0.2575 $\pm$ 0.0028  &0.0693 $\pm$ 0.0025 \\
\boldmath $\gamma$ & \textbf{2.018 $\pm$ 0.057} & \textbf{1.952 $\pm$ 0.098} & \textbf{1.8769 $\pm$ 0.0040} 
 & \textbf{2.346 $\pm$ 0.013} \\
$s_h$       &18.7 $\pm$ 3.8   &22 $\pm$ 15      &0.00903 $\pm$ 0.00058&0.383 $\pm$ 0.041 \\
$K$         &0              &41.2$ \pm$ 6.5   &0                  &33.5 $\pm$ 5.9 \\\mr
DOF         &   232         &231            &232                &231 \\
$\chi^2/$DOF&    1.37       &1.15           &1.14               &1.11  \\\br 
\endfulltable


\subsection{Discussion and Partial Conclusions}\label{subsec:2.5}

\begin{figure}
\includegraphics[width=15cm,height=7.5cm]{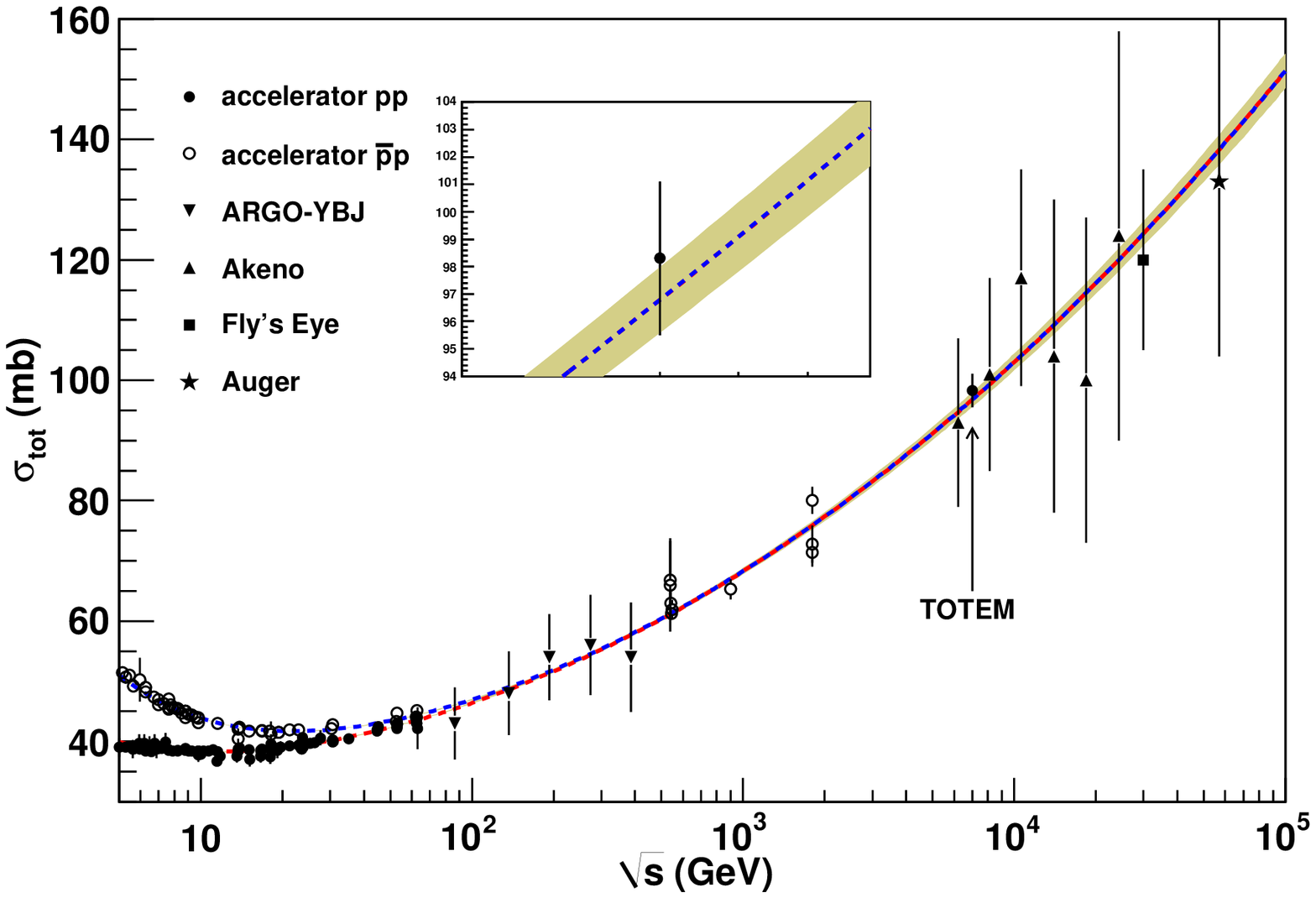}
\includegraphics[width=15cm,height=7.5cm]{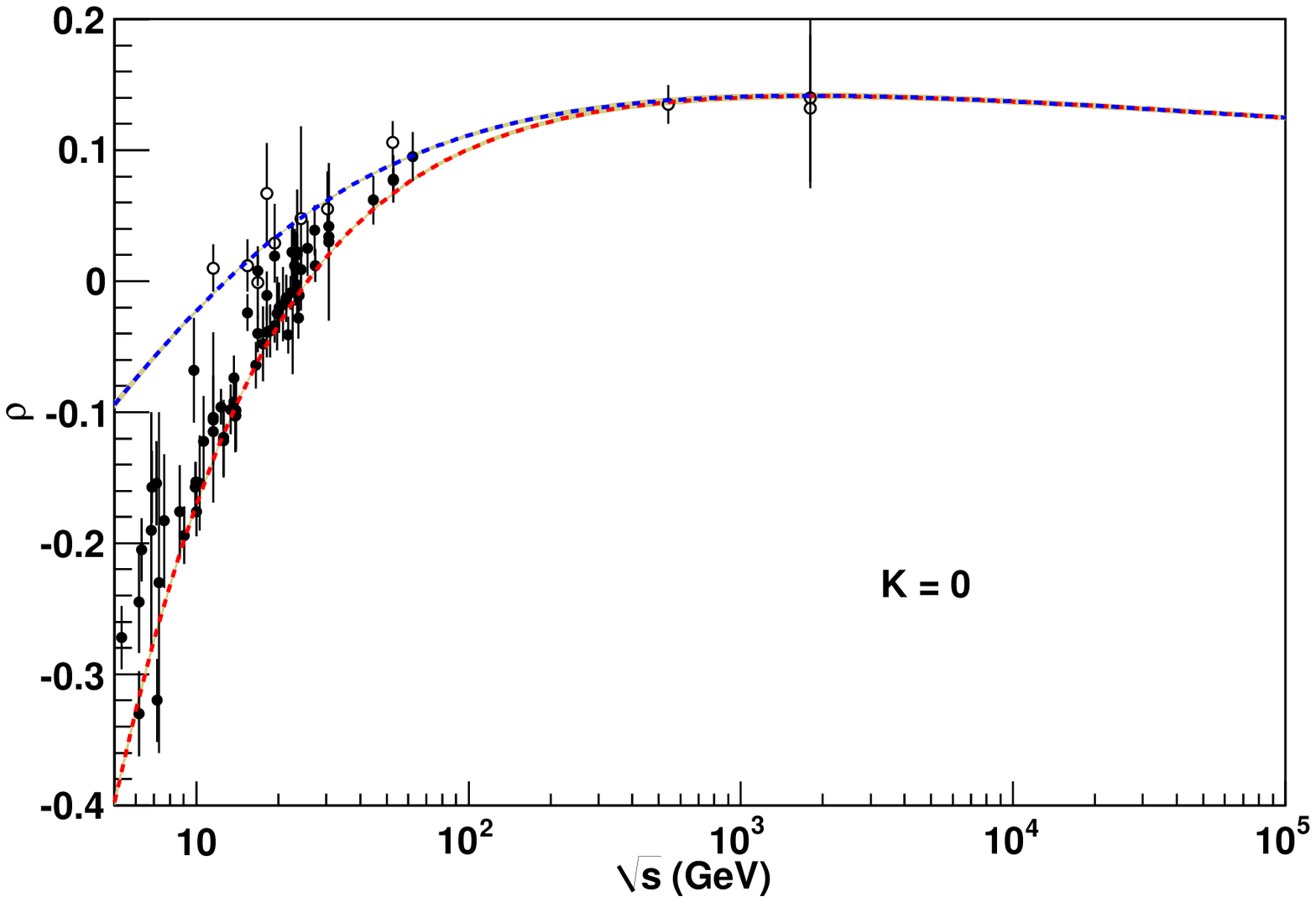}
\includegraphics[width=15cm,height=7.5cm]{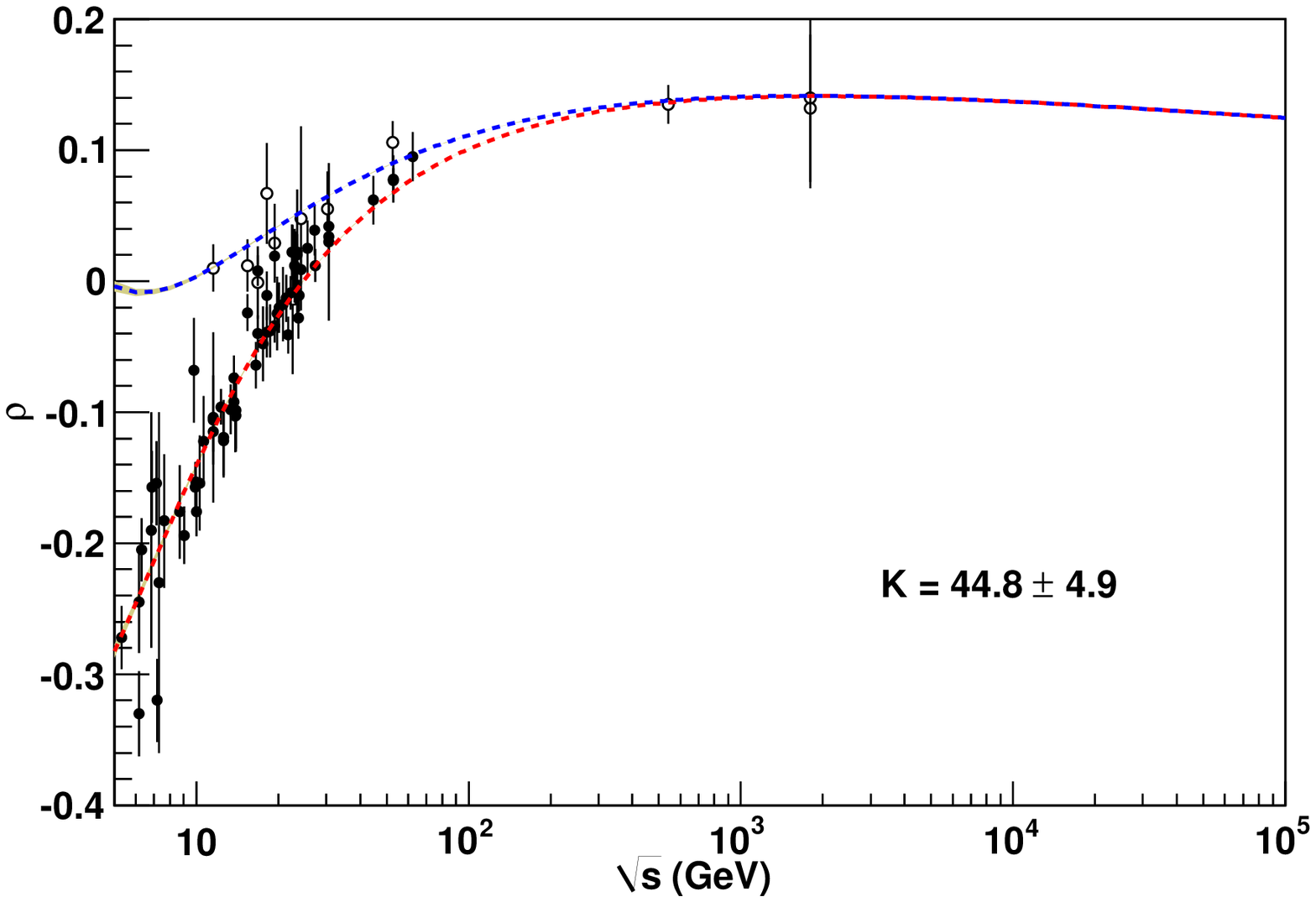}
\caption{Individual fit to $\sigma_{tot}$ data through parametrization
(3-4) from fit 1 and predictions for $\rho(s)$ with $K$ = 0 and $K$ as a free
fit parameter (\tref{tab:1}).}
\label{fig:1}
\end{figure}

\begin{figure}
\includegraphics[width=15cm,height=7.5cm]{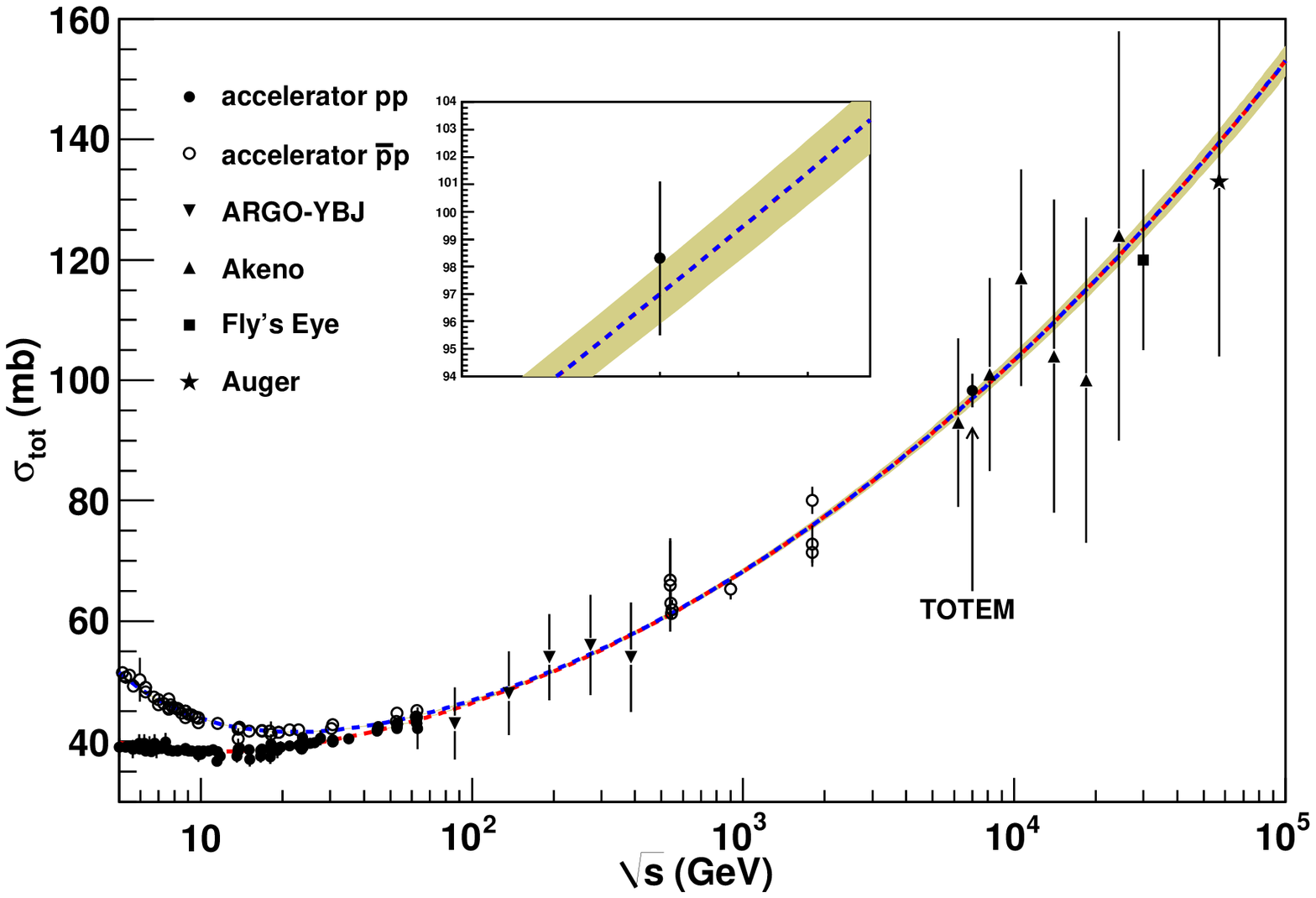}
\includegraphics[width=15cm,height=7.5cm]{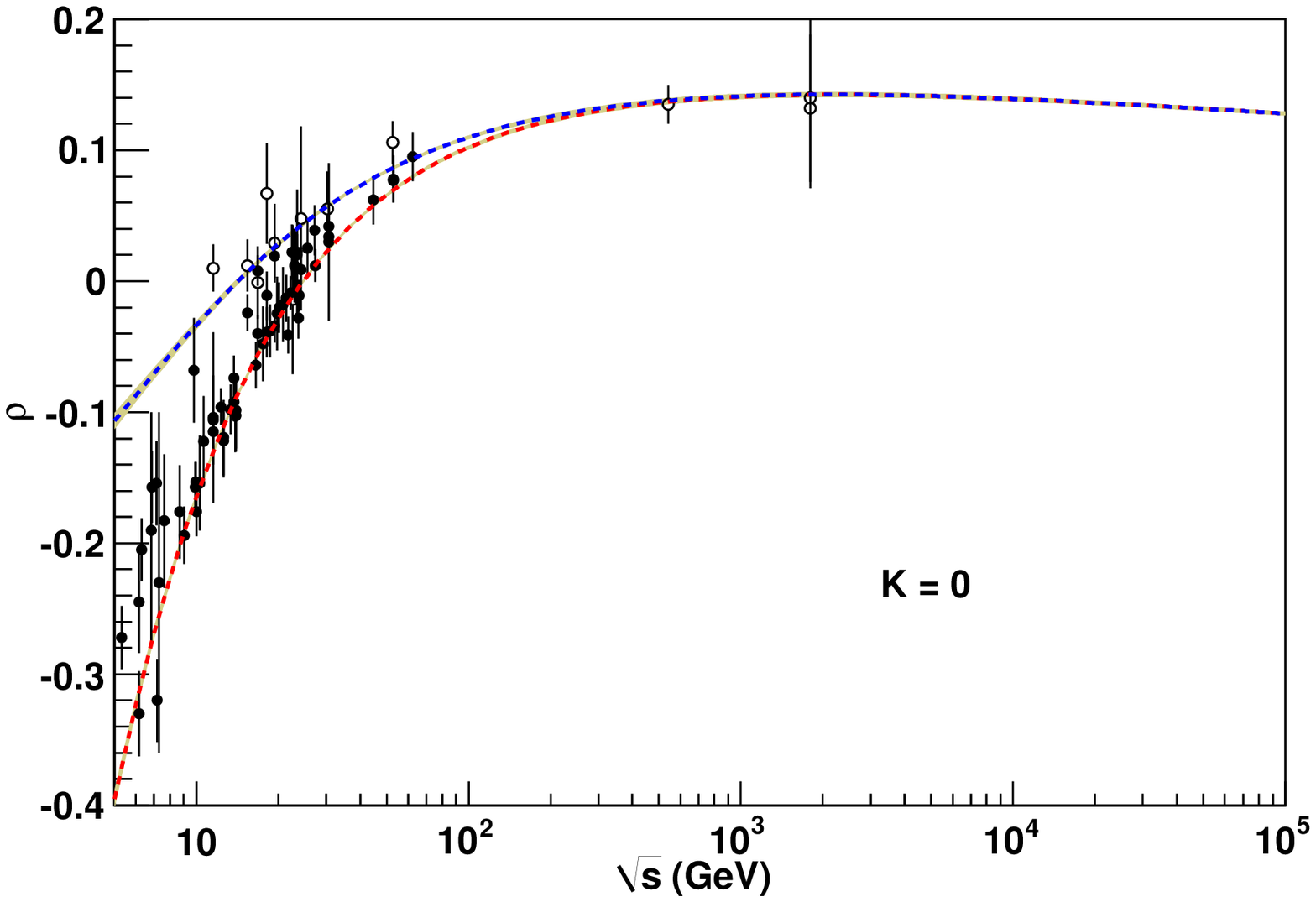}
\includegraphics[width=15cm,height=7.5cm]{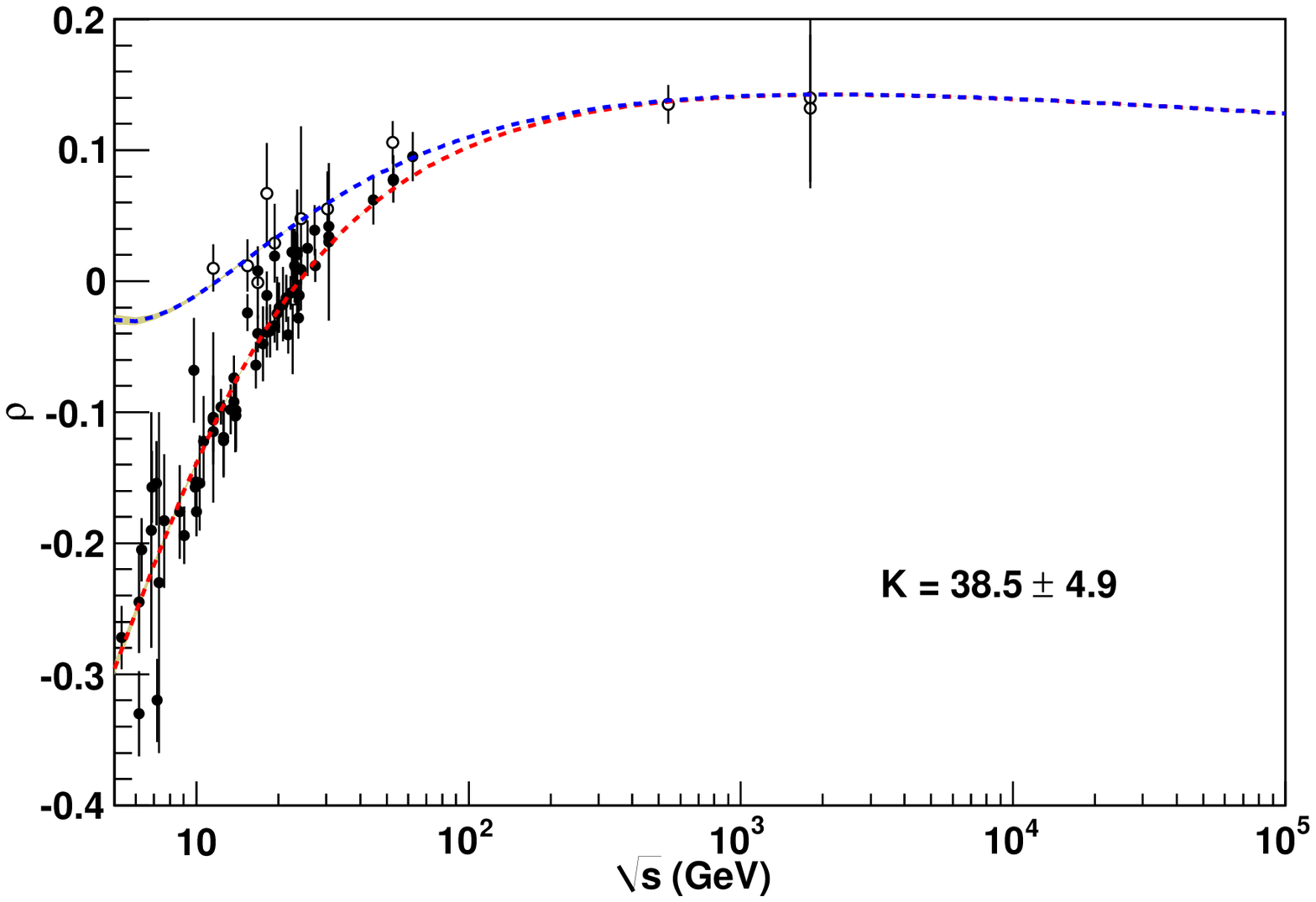}
\caption{Individual fit to $\sigma_{tot}$ data through parametrization
(3-4) from fit 2 and predictions for $\rho(s)$ with $K$ = 0 and $K$ as a free
fit parameter (\tref{tab:1}).}
\label{fig:2}
\end{figure}

\begin{figure}
\includegraphics[width=15cm,height=11cm]{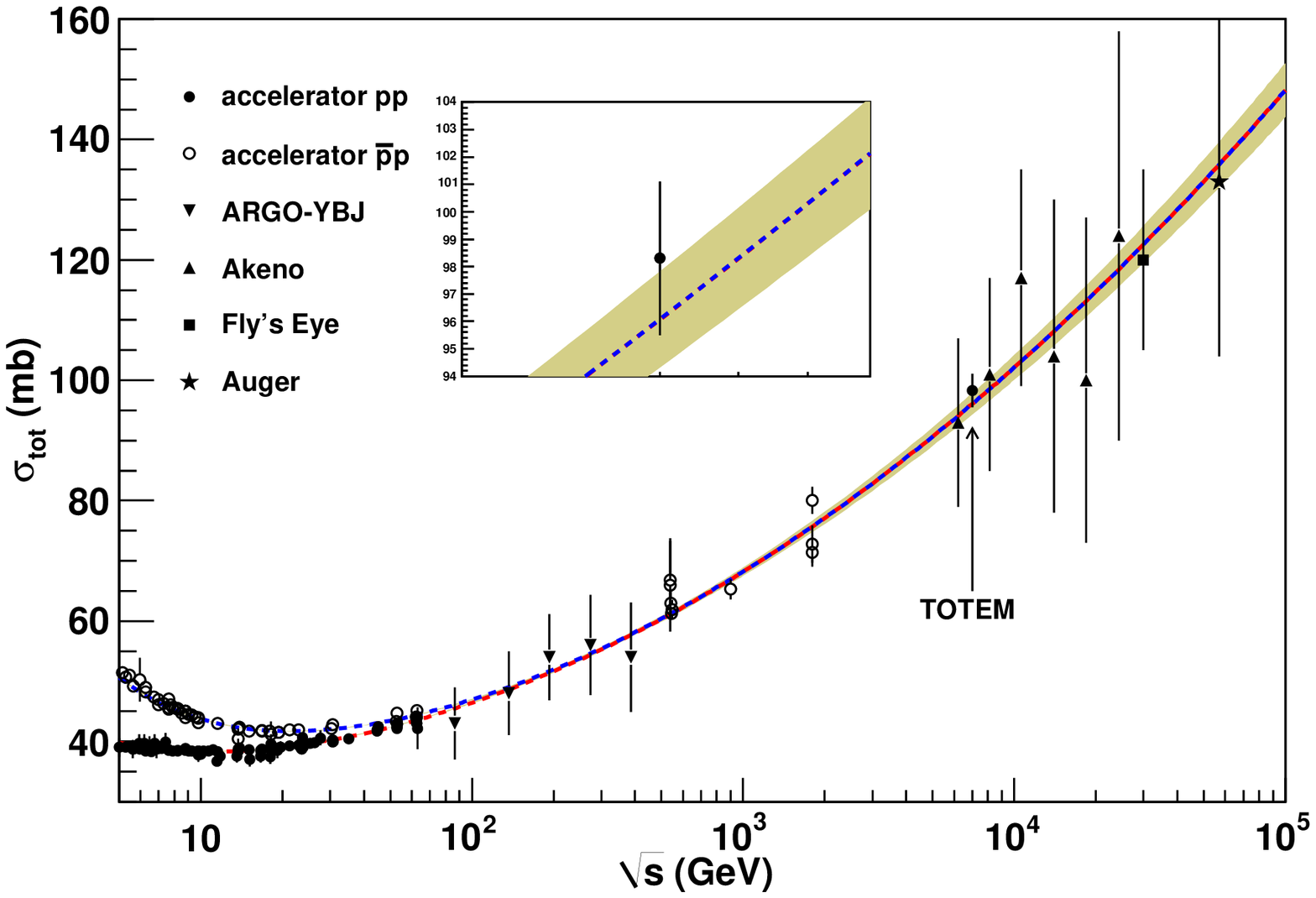}
\includegraphics[width=15cm,height=11cm]{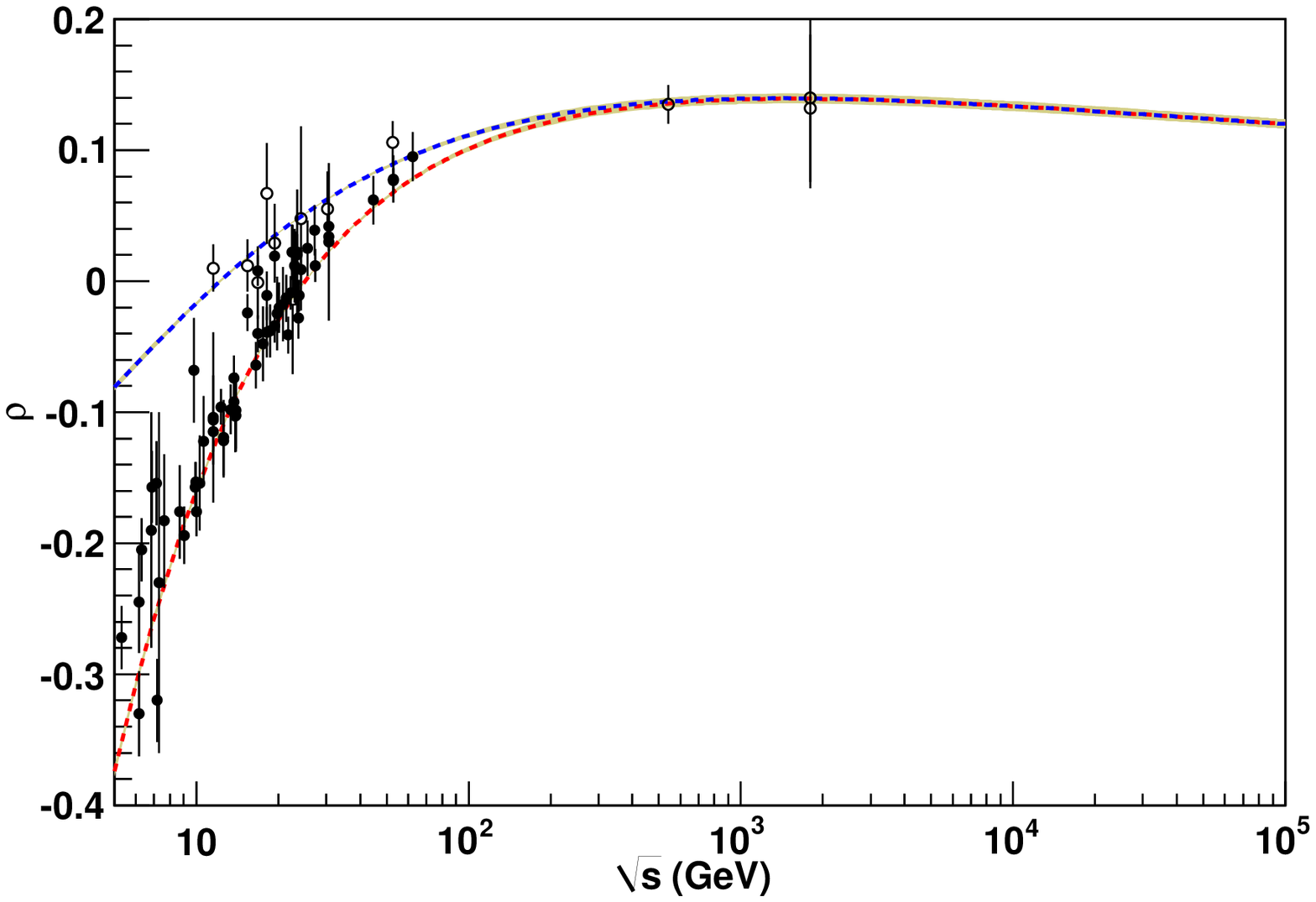} 
\caption{Simultaneous fits to $\sigma_{tot}$ and $\rho$ data from fit 1 and
$K$ = 0 (\tref{tab:2}).}
\label{fig:3}
\end{figure}

\begin{figure}
\includegraphics[width=15cm,height=11cm]{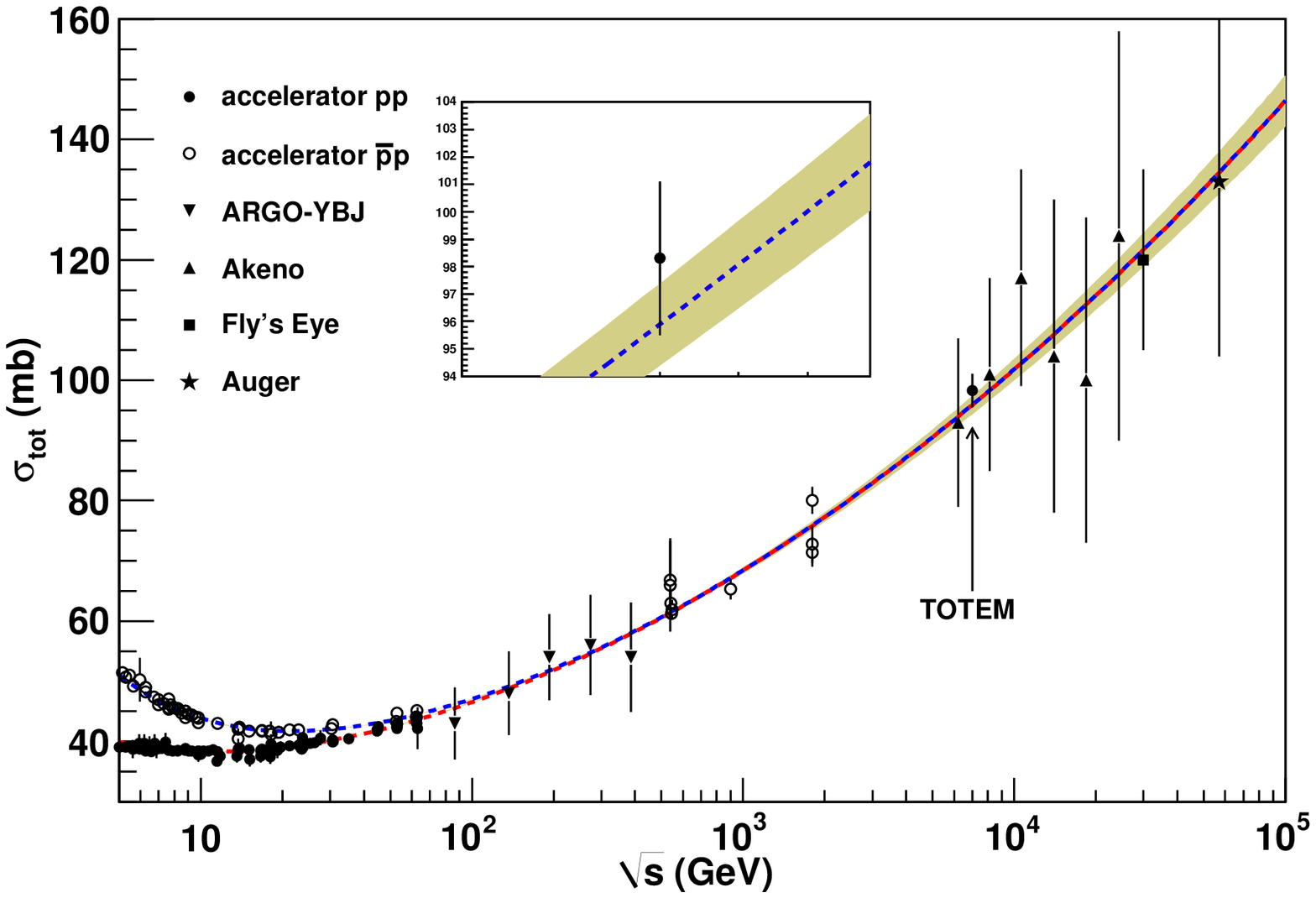}
\includegraphics[width=15cm,height=11cm]{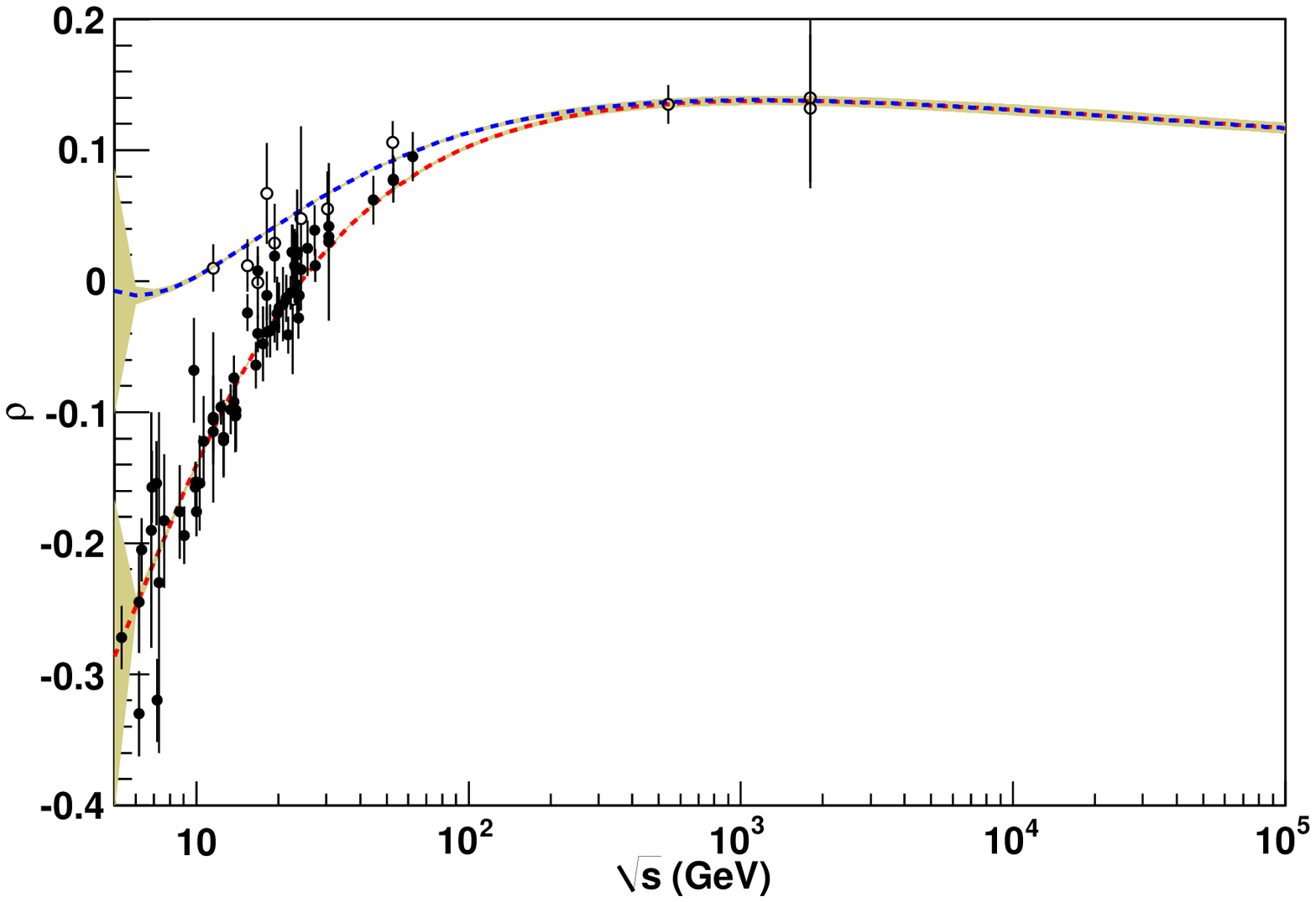} 
\caption{Simultaneous fits to $\sigma_{tot}$ and $\rho$ data from fit 1 and
$K$ as a free parameter (\tref{tab:2}).}
\label{fig:4}
\end{figure}

\begin{figure}
\includegraphics[width=15cm,height=11cm]{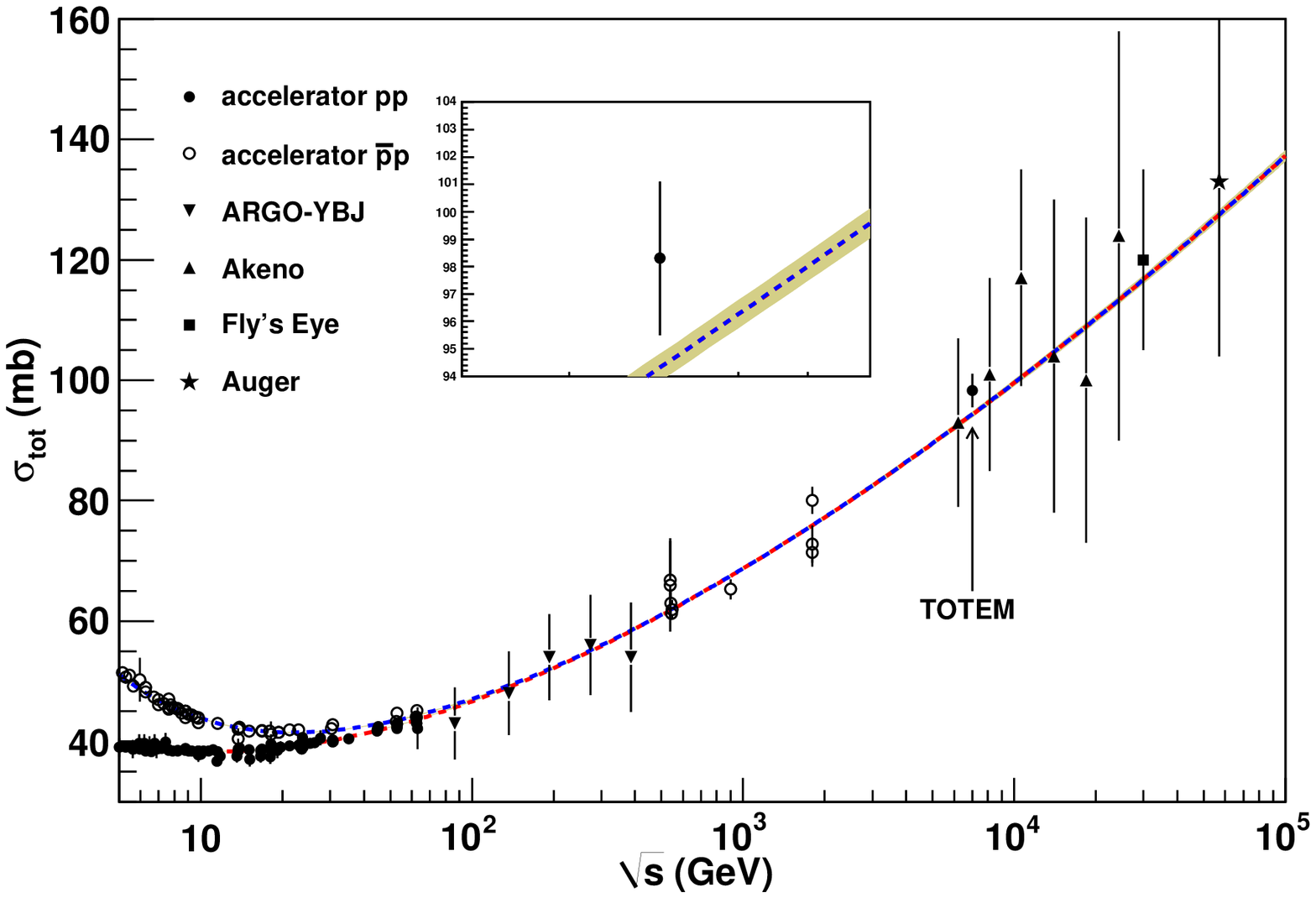}
\includegraphics[width=15cm,height=11cm]{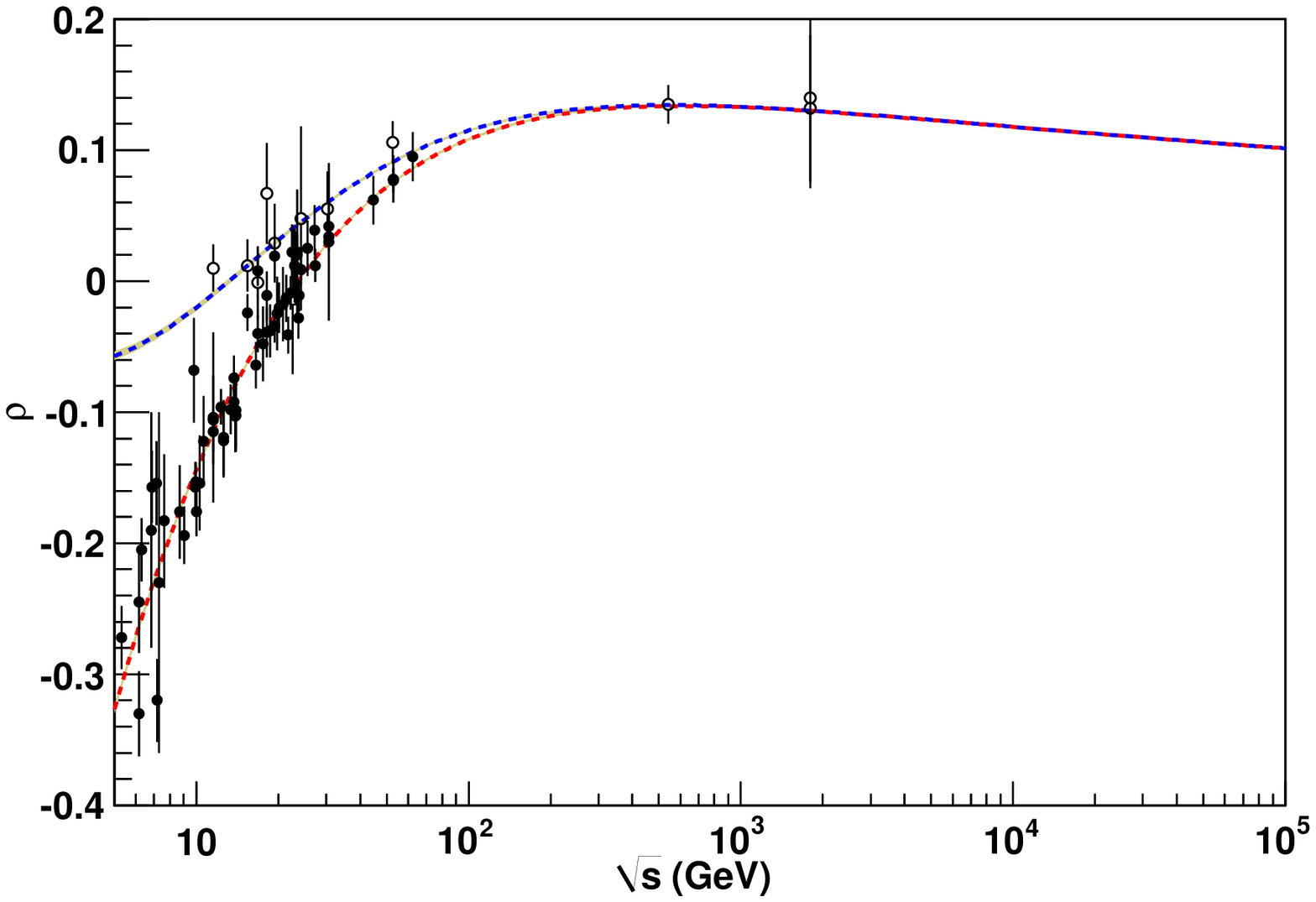} 
\caption{Simultaneous fits to $\sigma_{tot}$ and $\rho$ data from fit 2 and
$K$ = 0 (\tref{tab:2}).}
\label{fig:5}
\end{figure}

\begin{figure}
\includegraphics[width=15cm,height=11cm]{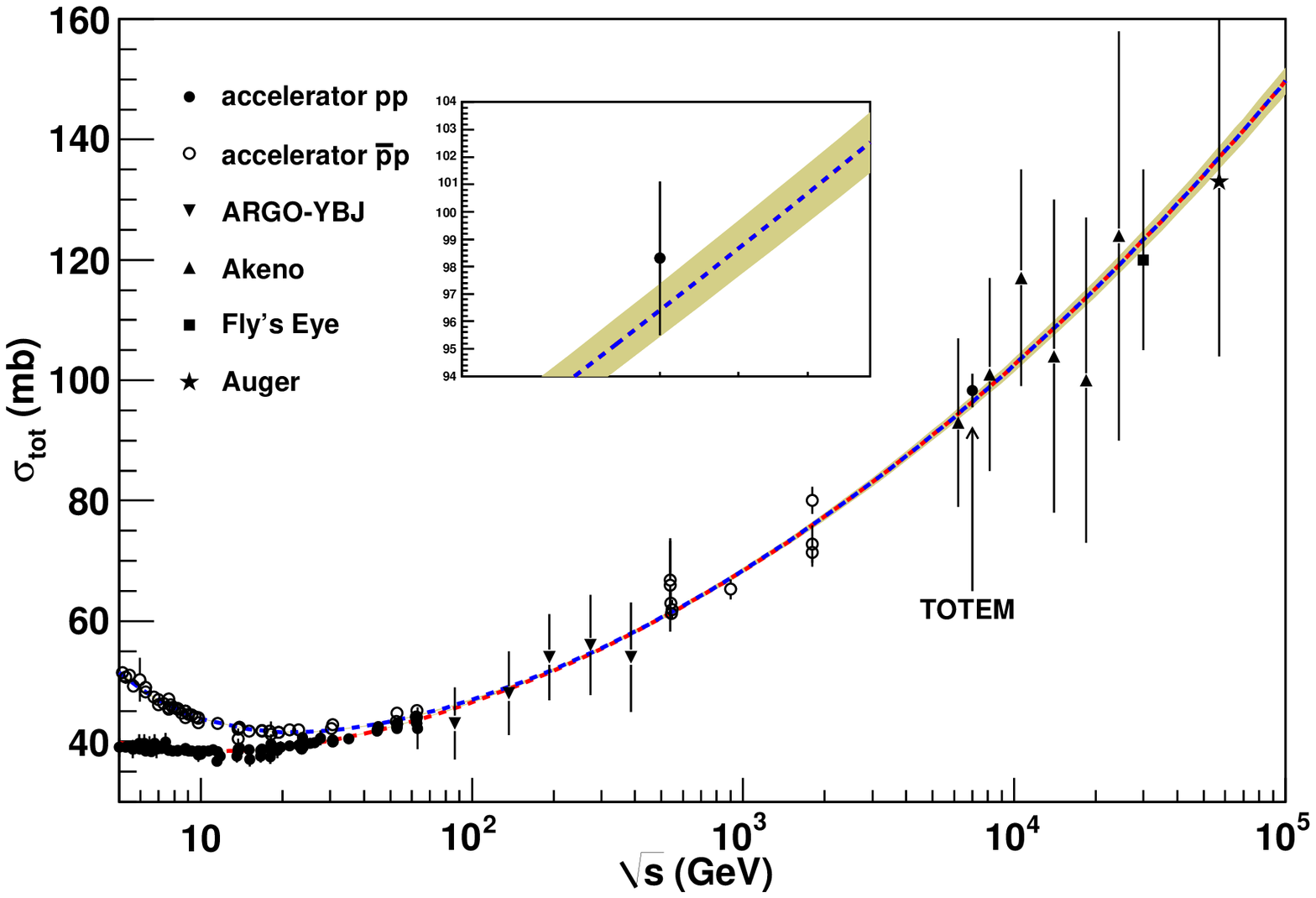}
\includegraphics[width=15cm,height=11cm]{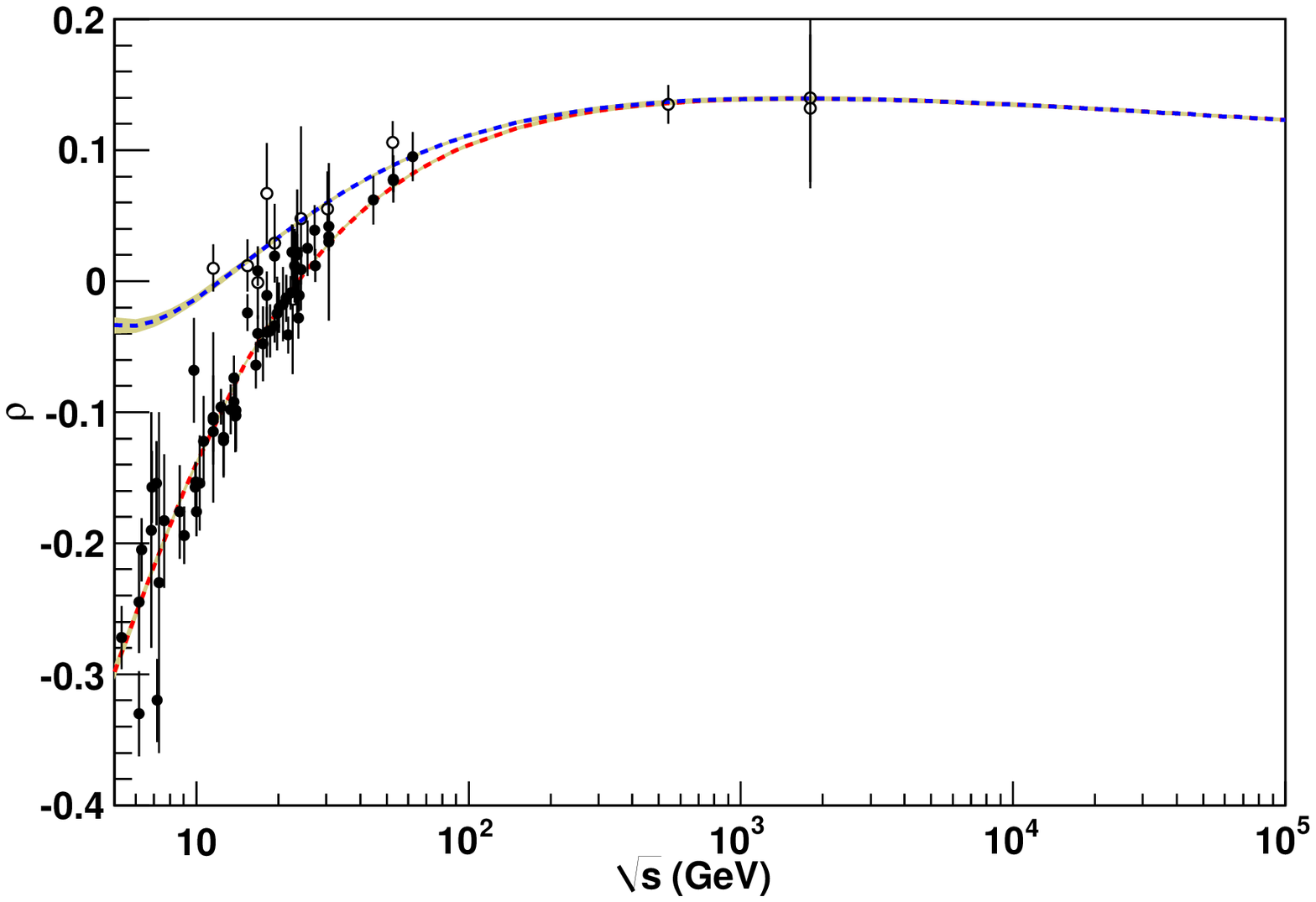} 
\caption{Simultaneous fits to $\sigma_{tot}$ and $\rho$ data from fit 2 and
$K$ as a free parameter (\tref{tab:2}).}
\label{fig:6}
\end{figure}

Beyond the distinct start values for the parameters in Fits 1 and 2, an essential difference concerns
the reggeon intercepts, which are fixed to $b_1 = b_2 =$ 0.5 in the former case and  free in the latter.
Based on this feature and the results presented in tables \ref{tab:1} and \ref{tab:2} and figures \ref{fig:1} - \ref{fig:6}, we have the comments
that follow, including some partial conclusions.

\vspace{0.5cm}

\noindent
$\bullet$ 
Individual Fits to Total Cross Section Data and Predictions for the Rho Parameter
(\tref{tab:1}, figures \ref{fig:1} and \ref{fig:2})

\begin{enumerate}

\item[1.] Both fits indicate $\gamma > $ 2 with good statistical confidence (\tref{tab:1}):

fit 1: $\gamma \sim$ 2.21 $\pm$ 0.02 and $\chi^2$/DOF $\sim$ 0.97,

fit 2: $\gamma \sim$ 2.42 $\pm$ 0.02 and $\chi^2$/DOF $\sim$ 0.93. 

\item[2.] The TOTEM result at 7 TeV is in agreement with the fits 
(fit 1 and fit 2) within uncertainties (figures \ref{fig:1} and \ref{fig:2}).

\item[3.] In the predictions for the $\rho$ parameter, the subtraction constant affects
the results in the low energy region, $\sqrt{s} \lesssim $ 20 GeV, essentially a change
of curvature in $\rho^{\bar{p}p}$ around 10 GeV, in the case of $K$ free (figures \ref{fig:1} and \ref{fig:2}).

\item[4.]  In all cases, the global description of the $\rho$ data is quite good and,
as expected, the best statistical results are obtained with $K$ as a free fit parameter
(\tref{tab:1}).

\item[5.] All the results with fit 2 corroborate those obtained in the FMS analysis.

\end{enumerate}

\noindent
$\bullet$
Simultaneous Fits to Total Cross Section and Rho Data
(\tref{tab:2} and figures \ref{fig:3} - \ref{fig:6})

\begin{enumerate}

\item[1.] Compared with individual fits to $\sigma_{tot}$ data, 
the inclusion of $\rho$ data in simultaneous fits leads to a decrease
in the value of the exponent $\gamma$ in all cases analyzed
(tables \ref{tab:1} and \ref{tab:2}):

fit 1: $\gamma \sim$ 2.21 $\Rightarrow$ $\gamma \sim$ 2.01 ($K$ = 0) and $\gamma \sim$ 1.95 ($K$ free)

fit 2: $\gamma \sim$ 2.42 $\Rightarrow$ $\gamma \sim$ 1.88 ($K$ = 0) and $\gamma \sim$ 2.35 ($K$ free)

\item[2.] The results with fit 1 are consistent with the log-squared bound (\tref{tab:2}):

$\gamma \sim$ 2.02 $\pm$ 0.06 for $K$ = 0 and $\gamma \sim$ 1.95 $\pm$ 0.10 for $K$ free.

The lower error bar in the TOTEM datum is consistent with 
the upper uncertainty region of the fit result
(figures \ref{fig:3} and \ref{fig:4}).

\item[3.] The result with fit 2 for $K$ = 0 is consistent with the log-squared bound,
$\gamma \sim$ 1.877 $\pm$ 0.0004,
but the fit uncertainty region does not reach the error bar of the TOTEM datum at 7 TeV (\fref{fig:5}).

\item[4.] The result with fit 2 for $K$ free is consistent with an increase of the total cross
section faster than the log-squared bound, 
$\gamma \sim$ 2.35 $\pm$ 0.01 and
the lower error bar of the TOTEM datum is in agreement with the fit 
result within uncertainties
(\fref{fig:6}).

\end{enumerate}

As commented before, we have included the results with $K$ = 0 only to illustrate
some aspects of the subtraction constant, outlined above. Since  the formal and consistent applicability of the DDR
here used demands the subtraction constant as a free fit parameter, we shall focus our partial
conclusions on this case only.

Concerning fit 1, we understand that fixing the reggeon intercepts to 0.5, 
results in a substantial constraint in the possible increase of the
total cross section, when the $\rho$ data is included in simultaneous fits. That is a consequence
of the correlations among all the fit parameters, in particular those involving
$K$, $s_h$ and $\gamma$. On the other hand, with fit 2, the reggeon
intercepts and strengths are free to control the low and intermediate energy data, leaving
the parameters associated with the constant and the  leading logarithmic contributions
adequately applied to the highest energy data. The correlation coefficients for both
fits with $K$ as a free fit parameter are displayed in Appendix A, where the correlations
among $K$, $s_h$ and $\gamma$ are stressed.
Moreover, fit 2 results in a $\chi^2$/DOF closest to 1 and within the uncertainty regions
all the experimental information on $\sigma_{tot}$ and $\rho$ is well described (\fref{fig:6}).

In addition, note that for gamma real (not an integer) 
the logarithm term $\ln^{\gamma}(s/s_h)$ is not defined for 
$s < s_h$ and that our energy cutoff is fixed at $s_{min}$ = 
25 GeV$^2$. In the case of fit 1, $s_h$ = 22 $\pm$ 15 GeV$^2$ 
reaches the physical region (above the energy cutoff)
within the uncertainty, leading, therefore,  to a 
questionable result. That, however, is not the case 
with fit 2 because $s_h$ = 0.38 $\pm$ 0.04 GeV$^2$.

Based on these facts, we shall select as our representative result that  obtained with fit 2
in the simultaneous fits to $\sigma_{tot}$ and $\rho$ data in the case of $K$ as a free fit parameter
(fifth column in \tref{tab:2}). This result, as well as those obtained with individual fits
to $\sigma_{tot}$ data (fits 1 and 2)
support an increase of the total cross section faster than the log-squared bound.

\vspace{0.5cm}

\section{Fits to Elastic Cross Section Data}\label{sec:3}

From s-channel unitarity, an increase of the total cross section faster than log-squared
is expected to be, in some way, connected with a similar effect in the elastic and/or 
inelastic cross sections. At least that is what is suggested by the log-squared bound
of the total cross section, recently extended to the inelastic cross section, as recalled
in our introduction. Therefore, to go one step further, a point of interest is to test the 
applicability of the same parametrization
(3-4) to the elastic or inelastic cross section data.
In order to address this possibility we shall consider only the
\textit{elastic cross section} for the four reasons that follow.

\begin{enumerate}

\item[(i)] 
From the optical theorem the total cross section is connected with the \textit{elastic amplitude} 
in the forward direction.

\item[(ii)] Total and elastic cross sections can be directly measured or estimated without
the need of model assumptions. That however is not the case for the inelastic cross section
due to the single and double dissociative contributions. 

\item[(iii)] The least ambiguous way to estimate the inelastic cross section is from unitarity, 
namely $\sigma_{inel} = \sigma_{tot} - \sigma_{el}$  \cite{giulia1}, which suggests 
a more fundamental character for the last two quantities.

\item[(iv)] To select a statistically consistent ensemble of $\sigma_{inel}$ data 
for fit is not
an easy task due to both model-dependence in ``direct'' estimations and the
different data from different experiments and energies for $\sigma_{tot}$ and $\sigma_{el}$, if using unitarity.

\end{enumerate}

The experimental data on the elastic cross section from $pp$ and $\bar{p}p$ scattering
have been extracted from the PDG database \cite{pdgsite}, without any kind of data
selection or sieve procedure.
For the fits to the elastic cross section data with parametrization (3-4) we have used as
start values those obtained with our representative result: fit 2 and $K$ as a free fit parameter 
in the simultaneous fits to $\sigma_{tot}$ and $\rho$ data. 

\begin{figure}
\includegraphics[width=15cm,height=10cm]{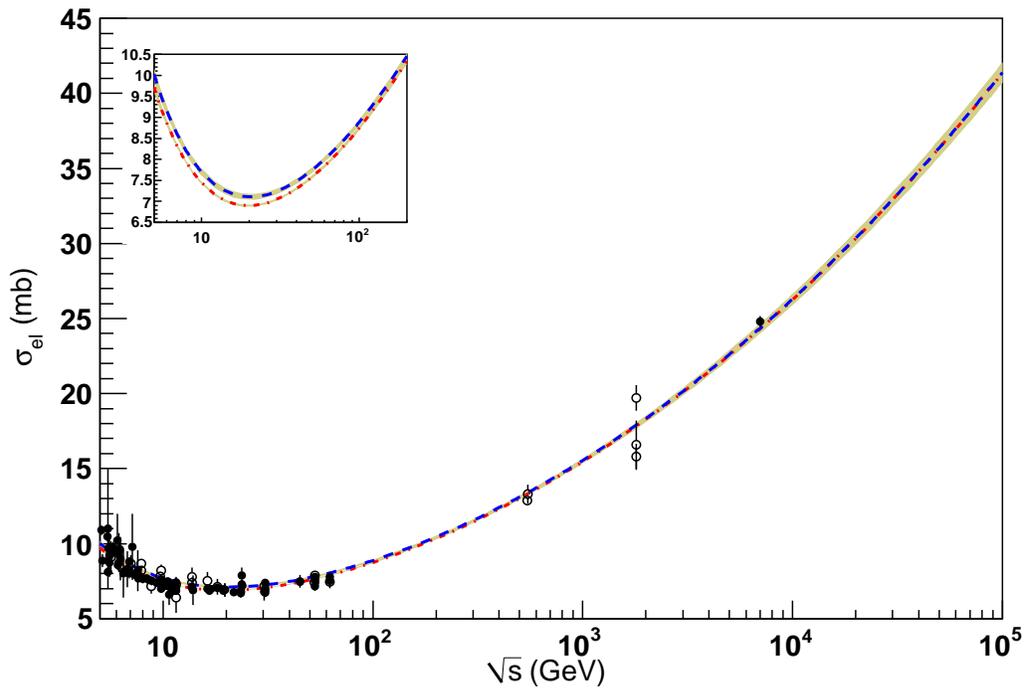} 
\caption{Fit result through parametrization (3-4) for the elastic cross section (\tref{tab:3})
and experimental data.}
\label{fig:7}
\end{figure}

In this case, a crucial point concerns the exponent $\gamma$ that controls the asymptotic
behavior of the cross sections. Given that we have obtained 
for the total cross section
$\gamma_{tot}$ = 2.346 $\pm$ 0.013,  a value different from this
for the elastic cross section,  $\gamma_{el}$, has no physical meaning,
due to direct violation of unitarity by the ratio $\sigma_{el}/\sigma_{tot}$
or the reduction of this ratio to zero at asymptotic energies.
Therefore we shall \textit{explore the possibility that the same exponent applies for the
elastic cross section} as well. Specifically, we fix this parameter to the central value
obtained for the total cross section, $\gamma_{el}$ = 2.346.
With the above assumptions and fit procedures, we obtain the results displayed in \tref{tab:3} and \fref{fig:7},
corresponding therefore to a good reproduction of the experimental data.
We shall discuss this result together with predictions for other physical quantities in the
next section.

\Table{\label{tab:3}Results of the fit to $\sigma_{el}$ data through parametrization (3-4),
with feedback values from fit 2, $K$ free (fifth column in \tref{tab:2}).
Same legend as \tref{tab:1}.} 
\br
$a_1$       & 30.7 $\pm$ 3.6 \\
$b_1$       & 0.551   $\pm$ 0.037  \\
$a_2$       & 0.236 $\pm$ 0.071 \\
$b_2$       & 0.134  $\pm$ 0.012  \\
$\alpha$    & 4.28 $\pm$ 0.14  \\
$\beta$     & 0.02358 $\pm$ 0.00054 \\
$\gamma$    & 2.346 (fixed)\\
$s_h$       & 0.978 $\pm$ 0.023   \\
\mr
DOF         &   97     \\
$\chi^2/$DOF&    1.62  \\
\br
\endTable

\section{Predictions for Other Physical Quantities}\label{sec:4}

Based on the empirical fits developed for $\sigma_{tot}(s)$, $\rho(s)$ and $\sigma_{el}(s)$, we present here the predictions
for several physical quantities, followed by a discussion on each result.

\subsection{Inelastic Cross Section and Ratios Elastic/Total and Inelastic/Total}\label{subsec:4.1}

\begin{figure}
\includegraphics[width=15cm,height=10cm]{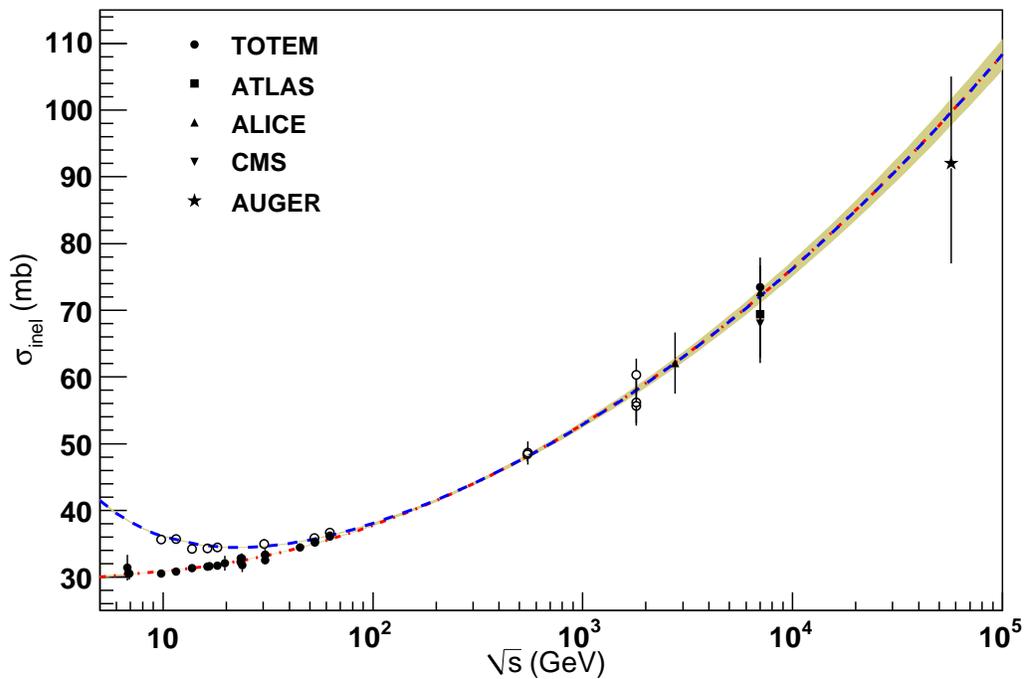} 
\caption{Predictions for the inelastic cross section and experimental data.}
\label{fig:8}
\end{figure}

\begin{figure}
\includegraphics[width=15cm,height=10cm]{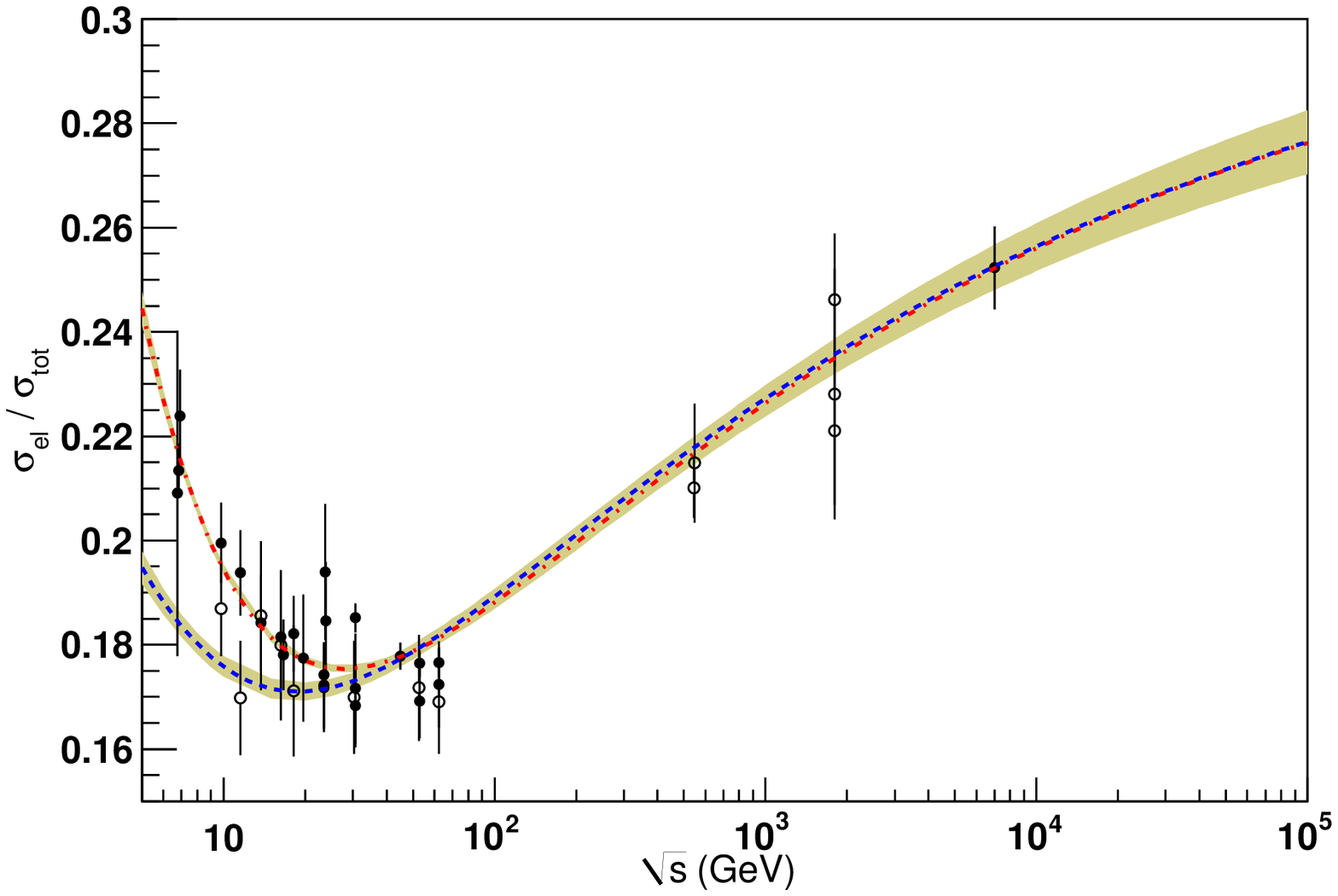} 
\caption{Predictions for the ratio between elastic and total cross section and experimental data.}
\label{fig:9}
\end{figure}

\begin{figure}
\includegraphics[width=15cm,height=10cm]{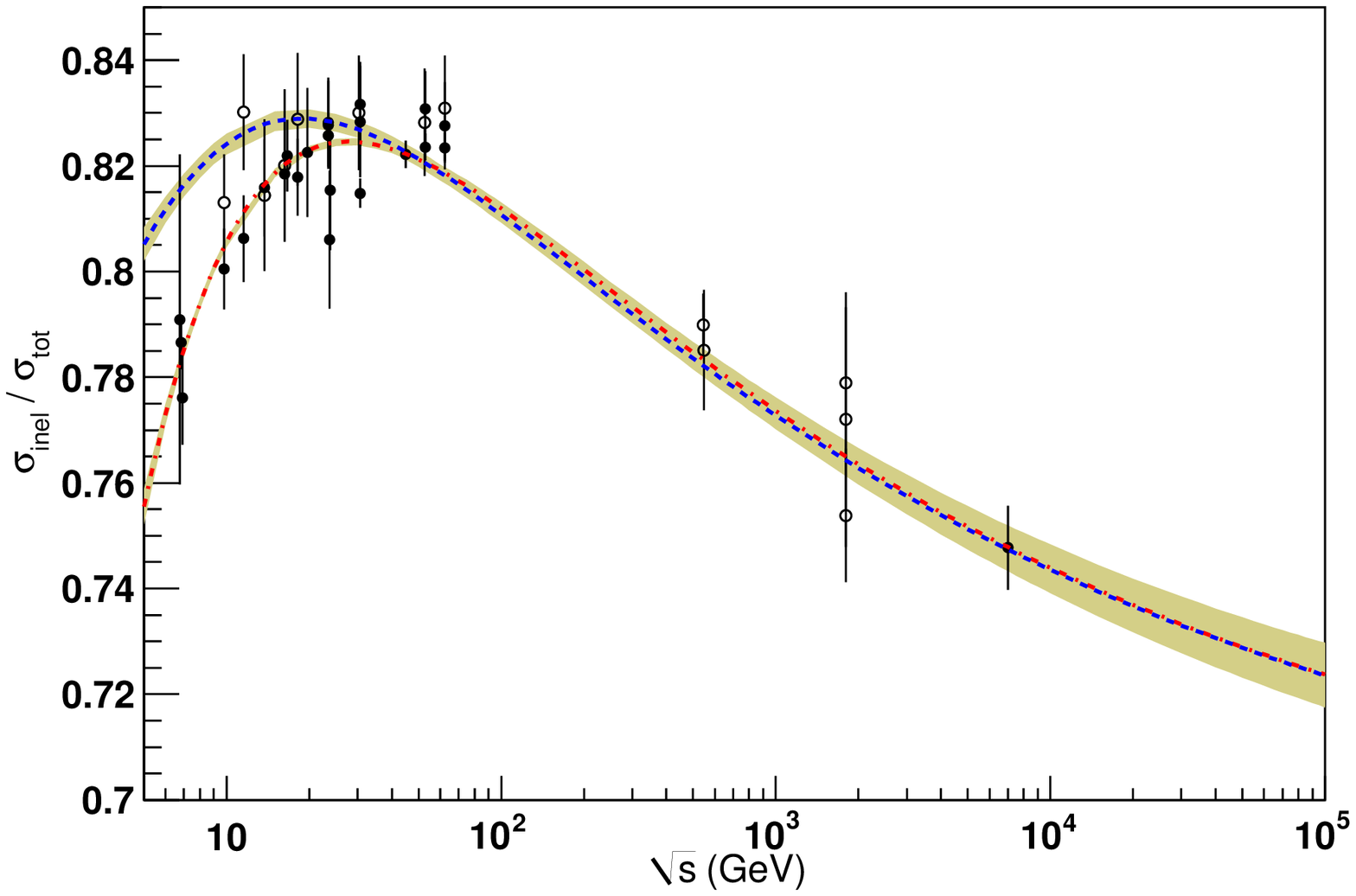} 
\caption{Predictions for the ratio between inelastic and total cross section and experimental data.}
\label{fig:10}
\end{figure}

From unitarity, by subtracting parametrization
(3-4) with the corresponding values of the parameters displayed in \tref{tab:2} for $\sigma_{tot}(s)$ and 
\tref{tab:3} for $\sigma_{el}(s)$,
we obtain our predictions for $\sigma_{inel}(s)$. The result is shown in \fref{fig:8}
together with the experimental data \cite{pdgsite,data1,data2,data3} and the uncertainty region. 
With this result and those for $\sigma_{tot}(s)$ and  $\sigma_{el}(s)$, we extract the corresponding predictions for the ratios $\sigma_{el}/\sigma_{tot}$ and 
$\sigma_{inel}/ \sigma_{tot}$, displayed in figures \ref{fig:9} and \ref{fig:10}, respectively.

\subsection{Optical Point, Elastic Slope and the Ratio with the Total Cross Section}\label{subsec:4.2}

In this section we first recall the definitions of the elastic slope $B$, the optical point
and an expression connecting
the ratios $\sigma_{el}/\sigma_{tot}$ and $\sigma_{tot}/B$.
After that we present our predictions.

In terms of the elastic amplitude, the differential cross section
is expressed by

\begin{eqnarray}
\frac{d\sigma}{dt}(s,t) = \frac{1}{16\pi s^2}\, |F(s,t)|^2, 
\end{eqnarray}
and the \textit{slope} of the elastic differential cross section in the forward direction
is defined as
\begin{eqnarray}
B(s,t=0) = \left[ \frac{d}{dt} \left(\ln \frac{d\sigma}{dt} \right) \right]_{t=0}.
\end{eqnarray}
From (1), (2) and (19), the \textit{optical point} is given by
\begin{eqnarray}
\left.\frac{d\sigma}{dt} \right|_{t=0} = 
\frac{\sigma_{tot}^2 [1 + \rho^2]}{16\pi}.
\end{eqnarray}
The integrated elastic cross section reads
\begin{eqnarray}
\sigma_{el}(s) = \int_{t_0}^{0}\,\frac{d\sigma}{dt}(s,t) dt,
\end{eqnarray}
where $t_0$ defines the physical (kinematic) region.

Concerning the differential cross section, experimental data indicate a sharp forward peak,
followed by a dip-bump or dip-shoulder structure above $\sim$ 0.5 GeV$^2$ (Tevatron, LHC).
Typically, these structures are located more than 5 decades below the optical point, equation (21).
These experimental facts are important in the determination of the integrated elastic
cross section, since in this case the differential cross section can effectively be
represented by an exponential fall off, simulated by a model-independent parametrization \cite{totemsigma},

\begin{eqnarray}
\frac{d\sigma}{dt} = \left.\frac{d\sigma}{dt} \right|_{t=0}\, e^{Bt},
\end{eqnarray}
with $B$ the (constant) forward slope. In that case, by assuming $t_0 \rightarrow - \infty$ 
in eq. (22),
the integrated elastic cross section is given by the approximate result

\begin{eqnarray}
\sigma_{el}(s) = \frac{1}{B(s)}\, \frac{\sigma_{tot}^2(s)}{16 \pi}[1 + \rho^2]
\nonumber
\end{eqnarray}
and therefore,
\begin{eqnarray}
\frac{\sigma_{tot}(s)}{B(s)} = \frac{16 \pi}{[1 + \rho^2]} \frac{\sigma_{el}(s)}{\sigma_{tot}(s)},
\end{eqnarray}
which is very close to the MacDowell-Martin  bound \cite{mac},
\begin{eqnarray}
\frac{\sigma_{tot}(s)}{B(s)} \leq 18 \pi \frac{\sigma_{el}(s)}{\sigma_{tot}(s)},
\end{eqnarray}
a result recently discussed in more detail in \cite{secb}.
For our purposes, the main ingredient in the result (24) is the possibility to
investigate the behavior of $\sigma_{tot}(s)/B(s)$ from the
information on the ratio $\sigma_{el}(s)/\sigma_{tot}(s)$ (see also \cite{secb}).

\begin{figure}
\includegraphics[width=15cm,height=10cm]{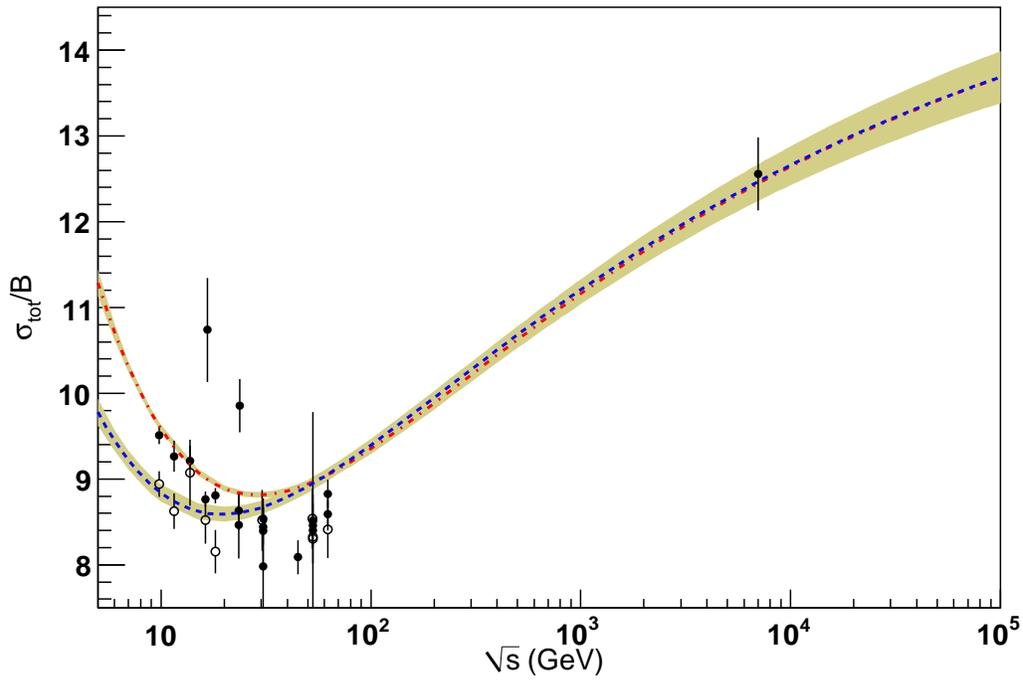} 
\caption{Predictions for the ratio between total cross section and the slope parameter and experimental data.}
\label{fig:11}
\end{figure}

\begin{figure}
\includegraphics[width=15cm,height=10cm]{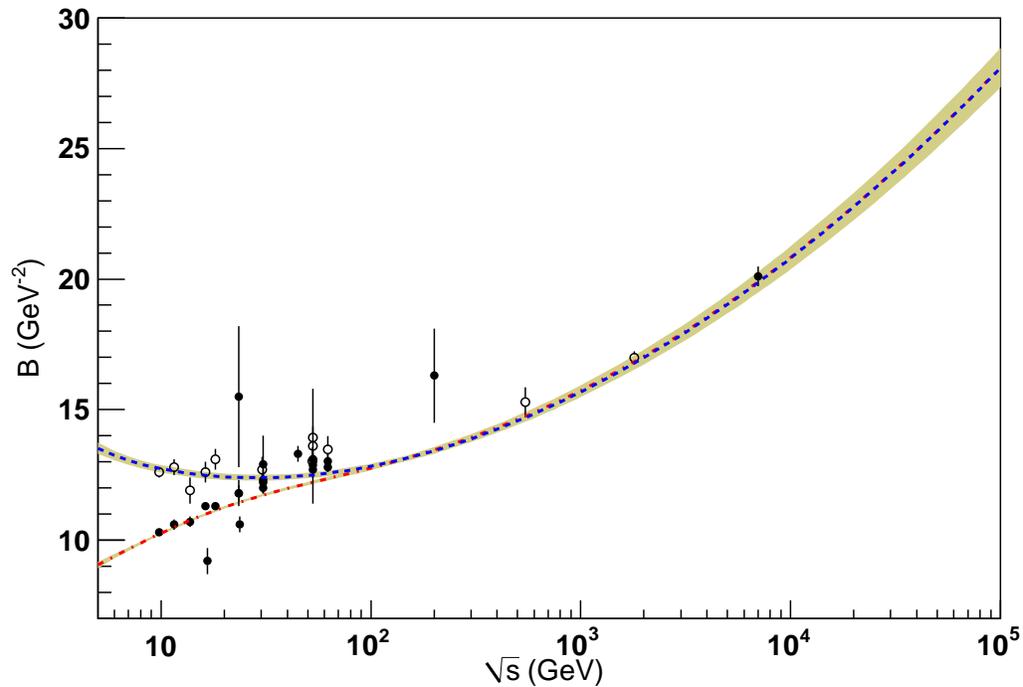} 
\caption{Predictions for the slope parameter and experimental data.}
\label{fig:12}
\end{figure}

With our empirical results for $\sigma_{tot}(s)$, $\rho(s)$ and $\sigma_{el}(s)$
we can predict the optical point (21), the ratio $\sigma_{tot}/B$ (24) and from that
the elastic slope $B$. The results for the last two quantities are shown in figures 
\ref{fig:11} and \ref{fig:12},
respectively (experimental information on the slope parameter B has
been compiled from the Durham database \cite{durham}).

The predictions for several quantities of interest at 7, 8, 14 and 57 TeV
are displayed in \tref{tab:4}, together with the TOTEM results at 7 TeV.


\begin{table}[h!]
\caption{\label{tab:4}TOTEM results at 7 TeV and predictions for several physical quantities associated with
the $pp$ elastic scattering at the LHC and Pierre Auger Observatory energies.\footnotesize}
{\footnotesize
\begin{tabular}{@{}llllll}
\br
Physical                   &  7 TeV  & 7 TeV & 8 TeV & 14 TeV & 57 TeV \\
quantity                   & TOTEM \cite{totemsigma} &  &  &  &  \\
\mr
$\sigma_{tot}$  (mb)        &98.3 $\pm$ 2.8&96.40 $\pm$ 0.97&98.7 $\pm$ 1.0&108.6 $\pm$ 1.2&137.0 $\pm$ 1.9     \\
$\rho$                      &  - &0.1362 $\pm$0.0016&0.1357 $\pm$ 0.0016& 0.1333 $\pm$ 0.0015&0.1261 $\pm$ 0.0013 \\
$\sigma_{el}$   (mb)        &24.8$\pm$ 1.2&24.31 $\pm$ 0.31&25.03 $\pm$ 0.33&28.18 $\pm$ 0.39&37.26 $\pm$ 0.59    \\
$\sigma_{inel}$ (mb)        &73.5 $\pm$ 3.1&72.1  $\pm$ 1.0&73.7 $\pm$ 1.1&80.4 $\pm$ 1.3& 99.7 $\pm$ 2.0       \\
$\sigma_{el}/\sigma_{tot}$  &0.252 $\pm$ 0.014&0.2522 $\pm$ 0.0041&0.2536 $\pm$  0.0042&0.2595 $\pm$ 0.0046&0.2720 $\pm$ 0.0057   \\
$\sigma_{inel}/\sigma_{tot}$&0.7477 $\pm$ 0.038&0.7478  $\pm$ 0.0041&0.7465 $\pm$ 0.0042& 0.7405 $\pm$ 0.0046&0.7280 $\pm$ 0.0057   \\
$\sigma_{tot}/B$            &12.56 $\pm$ 0.43&12.44 $\pm$ 0.21&12.51 $\pm$ 0.21&12.82 $\pm$ 0.23&13.46 $\pm$ 0.28      \\
$B$     (GeV$^{-2}$)        &20.10 $\pm$ 0.36&19.89 $\pm$  0.39&20.25 $\pm$ 0.40&21.76 $\pm$ 0.46&26.15 $\pm$ 0.66     \\
$d\sigma/dt|_{t=0}$         &504 $\pm$ 27&483.6 $\pm$ 9.7& 507 $\pm$ 10&613 $\pm$ 14&974 $\pm$ 27  \\
(mbGeV$^{-2}$) & & & & & \\
\br 
\end{tabular}
}
\end{table}

\subsection{Discussion}\label{subsec:4.3}

From \fref{fig:8} the prediction for $\sigma_{inel}(s)$ presents good agreement with all the
accelerator data and, within the uncertainty, also with the upper error bar of the 
Auger result at 57 TeV.
At 7 TeV it favors the TOTEM datum, a result obtained through unitarity and therefore
corresponding to a model-independent evaluation.

The results for the ratios of elastic/total and inelastic/total cross sections are displayed in
figures \ref{fig:9} and \ref{fig:10}. Within the uncertainty region, the predictions are also consistent with the TOTEM results at 7 TeV.
These results lead to the consequences that follows.

Asymptotically ($s \rightarrow \infty$), from parametrization (3-4) and by denoting the parameters associated with  $\sigma_{tot}$ and $\sigma_{el}$ fits by the corresponding indexes, we have

\begin{eqnarray}
\frac{\sigma_{el}}{\sigma_{tot}} \rightarrow \frac{\beta_{el}}{\beta_{tot}}
\qquad
\mathrm{and}
\qquad
\frac{\sigma_{inel}}{\sigma_{tot}} \rightarrow 1 - \frac{\beta_{el}}{\beta_{tot}}.
\nonumber
\end{eqnarray}
From tables \ref{tab:2} and \ref{tab:3} we obtain

\begin{eqnarray}
\frac{\sigma_{el}}{\sigma_{tot}} \rightarrow 0.340 \pm 0.015
\qquad
\mathrm{and}
\qquad
\frac{\sigma_{inel}}{\sigma_{tot}} \rightarrow  0.660 \pm 0.015,
\nonumber
\end{eqnarray}
which are consistent with the rational limits:
\begin{eqnarray}
\frac{\sigma_{el}}{\sigma_{tot}} \rightarrow \frac{1}{3}
\qquad
\mathrm{and}
\qquad
\frac{\sigma_{inel}}{\sigma_{tot}} \rightarrow \frac{2}{3},
\end{eqnarray}
indicating, therefore, a constant asymptotic value for the ratio $\sigma_{el}/\sigma_{tot}$ 
which lies below 1/2, the black-disc limit.

It is interesting to note that the above results corroborate recent arguments attributing the black-disc limit 
to a combination of the soft processes, namely elastic and diffractive. Specifically, denoting $\sigma_{diff}$ 
the single and double diffractive contributions, Grau-Pacetti-Pancheri-Srivastava have proposed the following asymptotic limit \cite{giulia2}

\begin{eqnarray}
\frac{\sigma_{el}}{\sigma_{tot}} + \frac{\sigma_{diff}}{\sigma_{tot}}
\rightarrow 
\frac{1}{2}
\qquad
\mathrm{as}
\qquad
s \rightarrow \infty.
\nonumber
\end{eqnarray}
This argument is based on the formulation of eikonal models \cite{giulia1,ll} and according to our results
it is expected:
\begin{eqnarray}
\frac{\sigma_{diff}}{\sigma_{tot}} \rightarrow \frac{1}{6}.
\nonumber
\end{eqnarray}

Another aspect concerns the recently proposed empirical parametrization for the ratio 
elastic/total cross section by Fagundes and Menon \cite{secb},
\begin{eqnarray}
\frac{\sigma_{el}}{\sigma_{tot}}(s) = A \tanh (\gamma_{1} + \gamma_{2} \ln s + \gamma_{3} \ln^{2} s ),
\end{eqnarray}
where $\gamma_i$, $i = $ 1, 2, 3 are free fit parameters and $A$ represents the
asymptotic limit. In this work the authors have considered  two extrema cases,
$A = 1/2$ (black-disc limit) and $A = 1$ (beyond the black-disc limit), with good descriptions
of the $pp$ data above 10 GeV. Our rational limit (26) implies  $A$ = 1/3 and therefore another 
scenario, lying asymptotically below the black-disc limit, as already pointed out.

The predictions for the ratio $\sigma_{tot}/B$ through equation (24) and for the elastic slope $B$
are displayed in figures \ref{fig:11} and \ref{fig:12}, respectively. We recall that, even in the forward
direction ($t$ = 0),  the determination of the
slope from the differential cross section data demands some interval in the momentum
transfer. That can explain some of the discrepant points in both figures since
the corresponding intervals from different experiments are not always the same.
Within the uncertainty regions, the predictions show good agreement with the experimental
information at high energies, specially with the TOTEM result at 7 TeV.
Asymptotically, from eq. (24) and since $\rho \rightarrow$ 0 (\sref{subsubsec:2.3.3}), our rational limit (26) gives
\begin{eqnarray}
\frac{\sigma_{tot}}{B} \rightarrow \frac{16\pi}{3}
\nonumber
\end{eqnarray}
and therefore, an asymptotic increase of the slope also faster than log-squared.
As commented before, this result represents a different scenario  from that
discussed in \cite{secb} and also in \cite{hadron2012}.

In \tref{tab:4} we display our numerical results (with uncertainties) for all the physical
quantities here discussed and at different energies of 
interest in present and future experiments. At 7 TeV, except for the $\rho$ parameter not determined by the
TOTEM Collaboration, the compatibility of our results with the experimental data is quite good, for all physical quantities.

At last, another aspect deserving some comments
concerns the predictions for the proton-proton total cross sections beyond
the (accelerator) energy region analyzed, namely
the estimations
of $\sigma_{tot}^{pp}$ from proton-air production cross section in cosmic-ray experiments.
As already commented, these points did not take part in our data reductions and have been displayed
only as qualitative illustrations. The estimations at the highest energies 
(Fly's Eye and Auger Collaborations)
depend on extrapolation 
from phenomenological models tested only in the accelerator energy region. This fact,
associated with the relative small cosmic-ray flux above $\sim$ 20 TeV, results in extremely large
uncertainties, as shown in our figures. Even though, the fit extrapolation
suggests good agreement with these points (\fref{fig:6}), in particular our numerical
prediction at 57 TeV (\tref{tab:4}) with the recent 
Auger Collaboration estimation for the total cross section,
$
\sigma_{tot}^{pp} = [ 133 \pm 13\, \mathrm{(stat)}^{+ 17}_{-20}\,\mathrm{(sys)} 
\pm 16\, \mathrm{(Glauber)}]\, \mathrm{mb}
$ \cite{auger}.
However, it is important to note that systematic and theoretical (Glauber) 
uncertainties are equally likely quantities, that is,  do not follow Gaussian 
distributions, as is the case of statistical uncertainties. That means 
the above central value is equally likely to lye in any place limited by 
the corresponding non-statistical uncertainties. For example, added in quadrature
the resulting uncertainty corresponds to $\approx$ 23 mb (around 17 \% of the central value).
This result contrasts with the small systematic uncertainty of the
TOTEM point a 7 TeV, namely 2.8 mb (around 2.8 \%).
Therefore, despite the suggestive global agreement, the large 
equally likely uncertainty 
may imply in different scenarios, as already discussed in [17]. 
We add that this observation puts limits on  
recent arguments by Block and Halzen in what concerns the description of the Auger Collaboration 
result quoted above \cite{bh2012}
(see also our critical remarks at the end of the next section).

\section{Conclusions and Critical Remarks}\label{sec:5}

In this work an amplitude analysis on the forward $pp$ and $\bar{p}p$ elastic scattering in the energy region
5 GeV - 7 TeV has been presented. The main point concerns the investigation of the log-squared
bound in the rise of the cross sections with the energy. To address this question,
an analytical parametrization introduced by Amaldi \textit{et al} \cite{amaldi},
with the exponent of the leading logarithm contribution as a free fit parameter, has been used.
Analytical connection between $\sigma_{tot}$ and  $\rho$ parameter
has been obtained by means of subtracted DDR.
The presentation has been divided in three parts: (1) individual and simultaneous
fits to $\sigma_{tot}$ and $\rho$ data, with focus on the role of the subtraction
constant; (2) extension of the parametrization to the elastic cross section
data with fixed exponent of the logarithm contribution; (3) predictions
for other physical quantities and the study of some consequences.

As we have shown and discussed in detail all analyzed and predicted data are quite well described.
The results indicate: (i) an increase of the hadronic cross sections faster than
the log-squared bound; (ii) asymptotic limits for the ratios elastic/total 
and inelastic/total cross sections consistent with 1/3 and 2/3, respectively.
Therefore, the violation of the bound does not imply in violation of unitarity,
as recently discussed by Azimov: the fast  rise
may correspond to a rapid high-energy increase of the scattering amplitude in
non-physical regions \cite{azimov1,azimov2}.

The possible rise of the total cross section faster than the
log-squared of $s$ may seems an unexpected result.
 However, that is  not the case
\textit{in what concerns the use of parametrization} (3-4), namely the possibility to treat
$\gamma$ as a free parameter and not to fix it to 2 (or 1). In fact, up to our knowledge,
before the FMS analysis, this parametrization was used in two works only,
the first one by Amaldi \textit{et al} in the seventies and the second by the UA4/2 
Collaboration in the nineties. Both analyses treated only $pp$ and $\bar{p}p$ elastic scattering
and simultaneous fits to $\sigma_{tot}$ and $\rho$ data. In the energy interval
5 GeV$ < \sqrt{s} \leq$ 62.5 GeV, Amaldi \textit{et al} obtained \cite{amaldi}
\begin{eqnarray}
\gamma = 2.10 \pm 0.10
\nonumber
\end{eqnarray}
and the UA4/2 Collaboration, with the data set extended up to 546 GeV  \cite{ua42},
\begin{eqnarray}
\gamma = 2.25^{+ 0.35}_ {- 0.31}.
\nonumber
\end{eqnarray}
These results, well known for a long time, indicate 
the possibility of a rise of the
total cross section faster than the log-squared bound.

In the FMS analysis two ensembles on $\sigma_{tot}$ data have been tested,
one with data up to $\sqrt{s}_{max}$ = 1.8 TeV and another one including the TOTEM result, 
namely $\sqrt{s}_{max}$ = 7 TeV. The novel aspect of the analysis consisted
in the evidence that with the former ensemble the $\gamma$ value is
consistent with 2, but above 2 with the latter one. In the case of 
$\sqrt{s}_{max}$ = 7 TeV,
several solutions have been obtained with different methods and variants \cite{fms11}:
\begin{eqnarray}\fl
\gamma \approx  2.10 \pm 0.03 \ (\mathrm{method\ 1},\ \mathrm{variant\ 1}),
 \qquad
\gamma \approx  2.27 \pm 0.04 \ (\mathrm{method\ 1},\ \mathrm{variant\ 3}), \nonumber \\
                                                                  \nonumber \\\fl
\gamma \approx  2.21 \pm 0.02 \ (\mathrm{method\ 2},\ \mathrm{variant\ 5}),
 \qquad
\gamma \approx  2.10 \pm 0.11 \ (\mathrm{method\ 2},\ \mathrm{variant\ 6}).
\nonumber
\end{eqnarray}
Although the fits had been restricted to $\sigma_{tot}$ data only, the use of 
DDR led to predictions for the $\rho$ parameter consistent 
with the experimental information, even with the largest $\gamma$ values \cite{fms11}.
Here, we have obtained with fit 2 to $\sigma_{tot}$ data
\begin{eqnarray}
\gamma \approx  2.42 \pm 0.02
\nonumber
\end{eqnarray}
and in the case of simultaneous fits to  $\sigma_{tot}$ and $\rho$ data,
\begin{eqnarray}
\gamma \approx  2.35 \pm 0.01.
\nonumber
\end{eqnarray}
Therefore, based on all the above results, we understand
as evident the fact that parametrization (3-4) applied to experimental data 
may lead to statistically consistent solutions with $\gamma >$ 2, possibly in the interval
2.2 - 2.4.

The leading contribution in parametrizations of the total cross section is expected to 
be associated with pomeron exchanges \cite{pred,land}.
For $\gamma$ = 1, the constant plus $\ln s$ terms correspond to a double pole at $J$ = 1
and for $\gamma$ = 2 a triple pole (expressed by $\ln^2 s$, $\ln s$ and the constant terms)
\cite{compete1,acm}. The case of non-integer exponent, with 0 $< \gamma <$ 2 corresponds
to a strong-coupling scenario \cite{gm,rmk} and a fractional power,
$\gamma$ = 3/2 (in general 1 $< \gamma <$ 2), is phenomenologically indicated
by the eikonal minijet model with soft gluon $k_t$-resummation
\cite{giulia1,giulia2,giulia3,giulia4}.
Presently, we do not have a physical interpretation for a leading 
logarithm contribution with the exponent $\gamma$ as a real
parameter greater than 2. Perhaps thinking about an effective exponent, representing
the sum of different contributions might not be so speculative.

Once associated with the leading contribution in parametrization (3-4), the $\gamma$ value
from data reductions is essentially determined by the experimental
information at the highest energy interval available. With $\sqrt{s}_{max}$ = 1.8 TeV
we are faced with the well known discrepant values of $\sigma_{tot}$ obtained
by the CDF, E710 and E811 Collaborations. This set of data gives no practical information
on the rise of the total cross section, except for a possible effective mean value 
in data reductions. It seems peculiar the fact that most phenomenological models
and even amplitude analyses present consistence with this \textit{imaginary} mean value.
On the other hand, compared with all the experimental data available, the TOTEM results
at 7 TeV are characterized by extremely small uncertainties, as can be easily seen in the 
figures. Our analysis has been essentially based on fits to total cross section and
elastic cross section, for which the uncertainties in the TOTEM data are around
2.8 \%  and 4.8 \%, respectively.
Therefore, in our data reductions, with the reduced $\chi^2$ as a test of goodness of fit,
these two points have played an essential role in the determination of the $\gamma$
parameter. In other words, we understand that our results indicating $\gamma >$ 2 are consequences
of the TOTEM results at 7 TeV, as also discussed in \cite{fms11}. 
The 2012 PDG analysis corroborates this conclusion since,
as already commented in our introduction, with the updated dataset the COMPETE log-squared parametrization does not
describe the TOTEM datum (Figure 46.10 in \cite{pdg12}); see also our recent
discussion \cite{replybhv2}, section III.A, specially Figure 1 in that paper.

It should be also noted that, contrasting with an
effective violation of the Froissart-Martin bound, a rise
of $\sigma_{tot}$ faster than log-squared of $s$ at the LHC
energy region, may be associated with some local
effect, so that, asymptotically the bound remains valid.
In that case, however, our asymptotic results, in particular
those concerning the ratios between elastic/inelastic
and total cross sections may have no meaning.

Therefore, based on the above critical comments, we understand that further measurements
on the total and elastic cross sections at the LHC energy region are
necessary before any firm conclusion can be drawn on the $\gamma$ value
and consequently on the results here presented and discussed. 
We also emphasize that
our results represent possible consistent statistical solutions for the behavior of the
total cross section, but do not correspond to unique solutions.
Despite these strong limitations, we hope this analysis and our previous work \cite{fms11}
might contribute with further interpretations and developments in the investigation of the rise of the
hadronic cross sections at high energies, a problem, in our opinion, still unsolved.

\textit{Note added}

After this analysis was completed, new high-precision
measurements of the total cross section at 7 and 8 TeV
have been reported by the TOTEM Collaboration 
\cite{totem12-1, totem12-2, totem12-3}.
An updated analysis, using the analytical parametrizations (1-5)
and including in the data set all these new measurements 
have been developed by Menon and Silva \cite{ms}. The obtained
results on the rise of the total cross section at high 
energy corroborate those presented here.

\vspace{0.3cm}


\ack

We are thankful to C. Dobrigkeit for a critical reading of the manuscript
and useful discussions.
Research supported by FAPESP 
(Contracts Nos. 11/15016-4, 11/00505-0, 09/50180-0).

\vspace{0.3cm}

\appendix
\section{Correlation Matrices}

\vspace{0.3cm}

In the MINUIT code the symmetric correlation matrix provides
a measure of the correlation between each pair of free parameter
through a coefficient with numerical limits $\pm$ 1 (full correlation)
and 0 (no correlation) \cite{bev,root}. The results from fits 1 and 2 are displayed in
\tref{tab:5}.

\fulltable{\label{tab:5}Correlation coefficients from the correlation matrices  associated with 
fit 1 and fit 2 results. The off-diagonal coefficients from fit 1 ($b_1 = b_2$ = 0.5 fixed)
 are displayed above the diagonal
of the table (not filled) and those from fit 2, below that diagonal.}
\br
 &          & \centre{9}{fit 1} \\
 &          & $a_1$  &  $b_1$ &  $a_2$ &  $b_2$ & $\alpha$ & $\beta$ & $\gamma$ &  $s_h$ & K     \\
\mr
\multirow{8}{*}{\ fit 2\ }
 &   $a_1$  &        &  $\:-$ &  0.063 & $\:-$  & -0.990   & -0.786  &  0.696  & -0.929  & 0.558  \\
 &   $b_1$  &  0.787 &        & $\:-$  & $\:-$  &  $\:-$   &  $\:-$  &  $\:-$  &  $\:-$  &  $\:-$ \\
 &   $a_2$  &  0.208 &  0.026 &        & $\:-$  & -0.009   & 0.063   & -0.073  & 0.027   & 0.271  \\
 &   $b_2$  &  0.176 &  0.020 &  0.976 &        &  $\:-$   &  $\:-$  &   $\:-$ &   $\:-$ & $\:-$  \\
 & $\alpha$ & -0.077 &  0.457 & -0.063 & -0.059 &          & 0.852   & -0.770  & 0.968   & -0.510  \\
 & $\beta$  &  0.061 & -0.062 & -0.005 & -0.008 & -0.139   &         & -0.987  & 0.954   & -0.304   \\
 & $\gamma$ & -0.139 & -0.131 &  0.041 &  0.039 &  0.178   & -0.816  &         &\textbf{-0.899}&\textbf{0.247}  \\
 &   $s_h$  & -0.252 & -0.129 &  0.063 &  0.062 &  0.557   &  0.364  &\textbf{0.120}  &        &\textbf{-0.424}   \\
 &   K      &  0.382 &  0.317 & -0.206 & -0.266 &  0.029   &  0.014  &\textbf{-0.031}  &\textbf{-0.057} &         \\
\br
\endfulltable

\noappendix

\end{document}